\documentclass[lettersize,journal]{IEEEtran}
\usepackage{amsmath,amsfonts}
\usepackage{algorithmic}
\usepackage{enumitem}
\usepackage[linesnumbered,ruled,vlined]{algorithm2e}
\usepackage{setspace}
\usepackage{array}
\usepackage{textcomp}
\usepackage{stfloats}
\usepackage{multirow,diagbox,tabularx,blindtext}
\usepackage{booktabs}
\usepackage{hyperref}
\usepackage{verbatim}
\usepackage{graphicx}
\usepackage{amssymb}
\usepackage{textcomp}
\usepackage{cite}
\usepackage{color}
\usepackage{pifont}
\usepackage{makecell}
\usepackage{threeparttable}
\usepackage{arydshln}
\usepackage[T1]{fontenc}
\usepackage[scaled=0.81]{beramono}

\newcommand{\PP}[1]{
\vspace{2px}
\noindent{\bf \IfEndWith{#1}{.}{#1}{#1.}}
}

\usepackage{subfigure}
\usepackage[normal]{caption}
\begin{document}

\begin{sloppypar}

\title{\huge LMAE4Eth: Generalizable and Robust Ethereum Fraud Detection by Exploring Transaction Semantics and Masked Graph Embedding} 

\author{Yifan Jia, Yanbin Wang*, Jianguo Sun*, Ye Tian, Peng Qian

\thanks{\emph{* Corresponding author: Yanbin Wang and Jianguo Sun.} }       

\thanks{Yifan Jia is with the Yantai Research Institute, Harbin Engineering University, Yantai 264000, China (e-mail:jiayf@hrbeu.edu.cn). Jianguo Sun and Ye Tian are with the Hangzhou Research Institute, Xidian University, Hangzhou 311231, China (e-mail:jgsun@xidian.edu.cn;messi.tianye@xidian.edu.cn). Yanbin Wang and Peng Qian are with the College of Computer Science, Zhejiang University, Hangzhou 310058, China (e-mail:wangyanbin15@mails.ucas.ac.cn;messi.qp711@gmail.com). }
}

\markboth{IEEE TRANSACTIONS ON INFORMATION FORENSICS AND SECURITY} 
{Jia \MakeLowercase{\textit{et al.}}: LMAE4Eth: Generalizable and Robust Ethereum Fraud Detection by Exploring Transaction Semantics and Masked Graph Embedding} 


\maketitle

\begin{abstract}
As Ethereum confronts increasingly sophisticated fraud threats, recent research seeks to improve fraud account detection by leveraging advanced pre-trained Transformer or self-supervised graph neural network. However, current Transformer-based methods rely on context-independent, numerical transaction sequences, failing to capture semantic of account transactions. Furthermore, the pervasive homogeneity in Ethereum transaction records renders it challenging to learn discriminative account embeddings. Moreover, current self-supervised graph learning methods primarily learn node representations through graph reconstruction, resulting in suboptimal performance for node-level tasks like fraud account detection, while these methods also encounter scalability challenges. 

To tackle these challenges, we propose LMAE4Eth, a multi-view learning framework that fuses transaction semantics, masked graph embedding, and expert knowledge. We first propose a transaction-token contrastive language model (TxCLM) that transforms context-independent numerical transaction records into logically cohesive linguistic representations, and leverages language modeling to learn transaction semantics. To clearly characterize the semantic differences between accounts, we also use a token-aware contrastive learning pre-training objective, which, together with the masked transaction model pre-training objective, learns high-expressive account representations. We then propose a masked account graph autoencoder (MAGAE) using generative self-supervised learning, which achieves superior node-level account detection by focusing on reconstructing account node features rather than graph structure. To enable MAGAE to scale for large-scale training, we propose to integrate layer-neighbor sampling into the graph, which reduces the number of sampled vertices by several times without compromising training quality. Additionally, we initialize the account nodes in the graph with expert-engineered features to inject empirical and statistical knowledge into the model. Finally, using a cross-attention fusion network, we unify the embeddings of TxCLM and MAGAE to leverage the benefits of both. We evaluate our method against 15 baseline approaches on three datasets. Experimental results show that our method outperforms the best baseline by over 10\% in F1-score on two of the datasets. Furthermore, we observe from three datasets that the proposed method demonstrates strong generalization ability compared to previous work. Our source code is avaliable at: \url{https://github.com/lmae4eth/LMAE4Eth}.
\end{abstract}


\section{Introduction}
 Blockchain technology has transformed many sectors by providing a secure and distributed way to record transactions \cite{intro_r1,chen2022zfw2}. Ethereum, launched in 2015, supports smart contracts and decentralized applications (dApps), enabling the creation of applications beyond simple financial transactions \cite{intro_r2,ma2025zfwdao}. It has become a key platform for innovations like decentralized finance (DeFi) and NFTs \cite{nlp_in_finance}. However, as Ethereum's value have grown, it has become a target for malicious actors seeking financial gain, with phishing and fraud emerging as major concerns within the ecosystem. The Chainalysis 2024 Crypto Crime Report notes that approximately USD \$23.2 billion in crypto-assets were transferred to illicit addresses, making up 0.34\% of total on-chain transactions \cite{chainalys2024}.

Reducing fraudulent activities on Ethereum and safeguarding users' assets is critical \cite{rabimba2021zfw1}, with detecting fraudulent accounts being an essential component of Ethereum's ecosystem security. Previous studies have shown that transaction records from Ethereum accounts (wallets) can be used to identify some fraudulent accounts. Some approaches use sequence models to encode transaction sequences, while others apply Graph Neural Networks (GNNs) and graph embeddings to learn account representations from transaction graphs. However, these methods have several limitations:

\begin{enumerate}
    \item \textbf{Insufficient attention to semantic information in transactions:} Transaction records are predominantly treated as decontextualized numerical data points during feature extraction, rendering these points context-independent. This leads to methods that fail to capture the contextual and intent, which may contain critical insights for fraud detection, when encoding these numerical sequences.
    \item \textbf{Inadequate differentiation of transactional information:} Ethereum transactions are highly homogeneous, which can blur the representation differences between transaction accounts. Existing methods do not consider the relationships and differences between samples during training, inadvertently introducing anisotropy in the generated embeddings. This makes it challenging to distinguish fraudulent accounts and transactions from legitimate ones.
    \item \textbf{Insufficient consideration of node-level information:} Recent graph-based detection methods have advanced through self-supervised graph representation learning \cite{ttagn,GNN_r7}, but they rely on graph reconstruction techniques that focus on structural information while neglecting node-level details. This approach has two key limitations. First, the lack of node-level information inherently leads to suboptimal performance in node-level tasks. Second, the structural features of the graph are highly dependent on the specific characteristics of the dataset, introducing potential biases and representation constraints.
    \item \textbf{Scalability challenges:} The size of the Ethereum account transaction graph grows exponentially with the scale of the dataset, resulting in significant scalability issues. As a result, previous approaches struggle to scale efficiently to large graphs and are prone to problems such as neighborhood explosion, overbalancing, or excessive compression.

    \item \textbf{Narrow focus in transaction analysis:} While transaction sequence encoding, graph-based representation, and expert features each provide unique insights, most existing techniques tend to concentrate on only one of these, overlooking the potential of combining all three viewpoints for a more complete model.
\end{enumerate}

To address emerging challenges, we propose LMAE4Eth, a dual-path unsupervised framework that integrates a pre-trained transaction language model with a self-supervised graph neural network. Specifically, we propose a transaction language model that transforms context-independent digital transaction records into linguistically interpretable representations, enabling Transformer to extract semantic insights. We further propose a token-aware self-supervised contrastive learning objective to learn fine-grained account differentiation details. Subsequently, we propose a masked account graph auto-encoder focusing on node feature representation rather than reconstructing the entire graph structure, which enhances fraud account node detection. In addition, this approach avoids neighborhood explosion by not using graph convolution or message passing mechanisms for node neighborhood aggregation. By employing cosine error for feature reconstruction, we also mitigate dimensional and vector norm influences to reduce over-smoothing. Given the dense nature of Ethereum transaction network account nodes, we futher propose a  masked graph auto-encoder with layer-neighbor sampling to facilitate seamless large-scale training without compromising embedding quality. We also initialize graph account nodes with expert-engineered features to incorporate global statistical information. Finally, we integrate embeddings from the transaction language model and the graph autoencoder with expert features through a cross-attention network.

Our main contributions include:
\begin{itemize}[left=0pt]
\item We propose a new Ethereum fraud account detection method, LMAE4Eth, which combines semantic encoding from pre-trained Transformers, self-supervised graph embeddings, and expert-engineered features. It outperforms previous best methods by over 10\% in F1-score across three public datasets.
\item We propose a transaction-token contrastive language model, by conceptualizing the representation of transaction sequences, that enables Transformer architectures to extract semantic intents embedded from account transaction data.
\item We leverage a token-aware contrastive learning pre-training objective to mitigate anisotropy in learned account embeddings caused by the high homogeneity of Ethereum account transactions.
\item We propose a masked account graph auto-encoder (MAGAE) that employs a masked generative self-supervised mechanism and cosine error for feature reconstruction, enabling superior node-level representations while circumventing neighborhood explosion and over-smoothing challenges.
\item To address computational scalability, we augment our MAGAE approach with a layer-neighbor sampling technique that uniformly samples a fixed number of neighbors per layer. This facilitates efficient training on large-scale graphs and reduces the edge budget by up to an order of magnitude compared to alternative sampling strategies.
\end{itemize}

\begin{figure*}[t]
    \centering
    \includegraphics[width=1\textwidth]{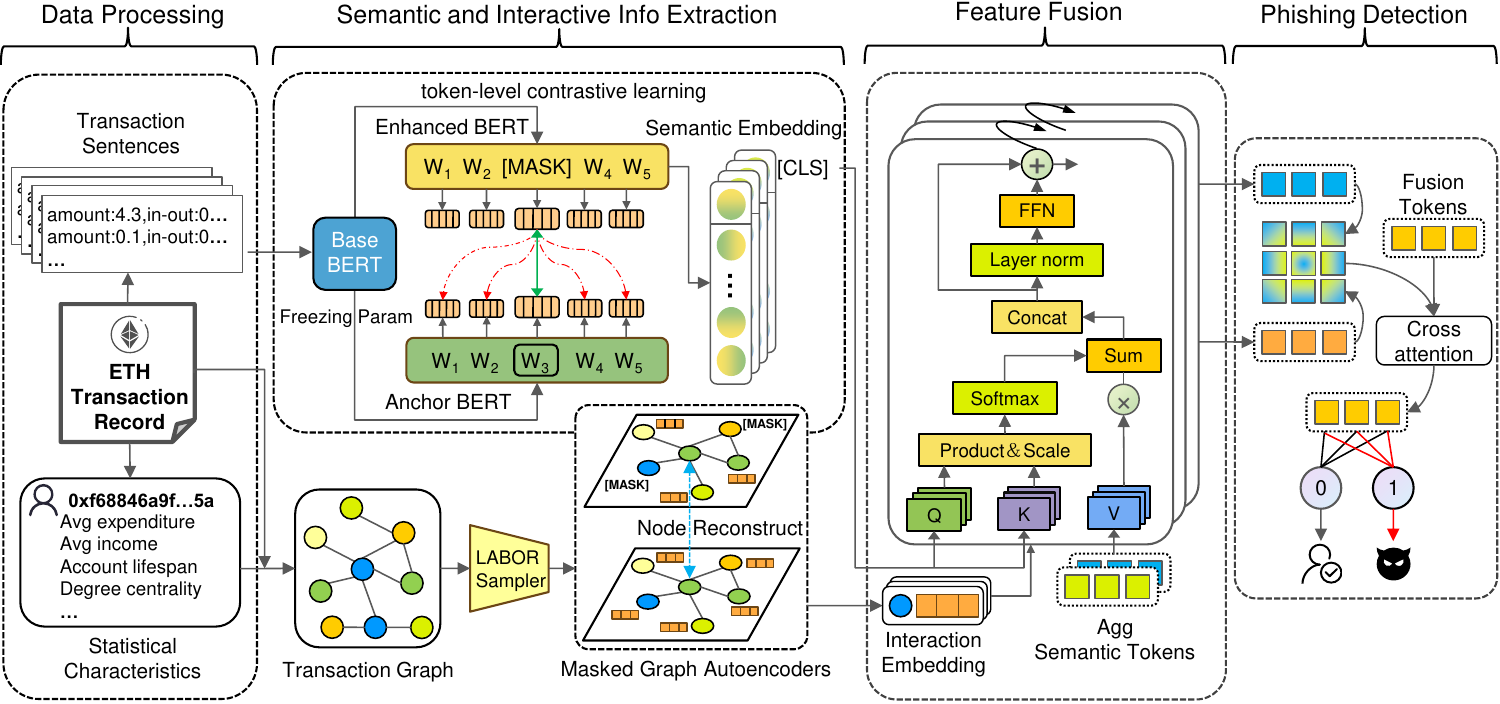}
    \caption{The framework of proposed Joint Transaction Language Model and Graph Representation Learning.}
    \label{fig:framwork}
\end{figure*}

\section{Backgroud and Related work}

\subsection{Ethereum Accounts and Transactions}

In the Ethereum, there are two primary types of accounts: Externally Owned Accounts (EOAs) and Contract Accounts. EOAs are controlled by private keys and can initiate transactions to transfer cryptocurrency or trigger smart contracts. These accounts are particularly relevant to phishing and fraud detection as they represent user-controlled activities\cite{bkg_zhang2024ethereum}. On the other hand, Contract Accounts are essentially smart contracts—self-executing programs deployed on the blockchain. Unlike EOAs, contract accounts cannot independently initiate transactions but can perform internal operations when triggered by an EOA\cite{bkg_zhou2021high,bkg_qin2022quantifying}.

Ethereum transactions are broadly categorized into two types:
\begin{itemize}[leftmargin=*]
    \item \textbf{External Transactions}: These transactions are initiated by EOAs, involving the transfer of cryptocurrency to other EOAs or triggering actions in contract accounts. They are central to our analysis as they directly reflect user activities and are more likely to reveal fraudulent behavior.
    \item \textbf{Internal Transactions}: These are transactions initiated by smart contracts within the blockchain. They are typically used for executing complex operations but do not provide direct insights into user-controlled activities.
\end{itemize}

Focusing on EOAs and external transactions is crucial for fraud detection, as they offer actionable insights into user behavior and potential phishing activities. An Ethereum transaction generally includes the following attributes:
\begin{itemize}[leftmargin=*]
    \item \textit{Sender}: The EOA initiating the transaction.
    \item \textit{Receiver}: The target EOA or contract account.
    \item \textit{Value}: The amount of cryptocurrency being transferred.
    \item \textit{Timestamp}: The time the transaction was recorded on the blockchain.
    \item \textit{Other Data}: Optional metadata used for contract execution or other supplementary information.
\end{itemize}

These attributes are critical for analyzing transactional patterns and detecting anomalies in blockchain activity.

\subsection{Graph-based Methods} 
The graph-based approach, by utilizing the transaction network between accounts, trains the model using graph embedding algorithms or Graph Neural Networks (GNN). For instance, Node2Vec has been applied to extract features from Ethereum transaction networks for fraudulent account detection \cite{node2vec,2vec_r1,2vec_r2}. Trans2Vec \cite{trans2vec}, built upon DeepWalk \cite{deepwalk}, divides representation learning into node, edge, and attribute levels to enhance classification performance.
Other studies have introduced behavior-based network embedding algorithms that utilize GCNs to classify transaction addresses into legitimate and fraudulent categories \cite{GNN_r3,GNN_r4}. A notable example, GAE-PDNA \cite{GNN_r7}, employs graph autoencoders to extract rich information from network nodes, demonstrating their effectiveness in capturing structural patterns. Additionally, subgraph segmentation methods have been proposed to construct transaction subgraphs for specified nodes. Improved graph classification algorithms are then applied to these subgraphs to achieve account classification \cite{GNN_r5,GNN_r6}.

Some methods extend graph-based approaches by adding extra information. TSGN \cite{tsgn} combines handcrafted features with Diffpool-based techniques to improve classification via mapping mechanisms in the transaction network. TTAGN \cite{ttagn} aggregates edge representations with Edge2Node, captures temporal sequences using LSTM, and concatenates statistical features for classification. GrabPhisher \cite{grabphisher} uses Node2Vec to initialize node vectors and time series analysis to detect fraudulent accounts.

However, Node2Vec, DeepWalk, and Trans2Vec, as shallow embedding methods, are not as effective in capturing the underlying complex node representations. While GCN and GAT are capable of scaling to deeper layers, they are prone to neighborhood explosion due to their use of graph convolutions or message-passing for neighborhood aggregation. Although recent approaches like GAE-PDNA attempt to address these limitations, they prioritize structural information, which results in suboptimal performance for node-level account detection tasks.

\subsection{Transformer-based Methods}
The graph-based approach faces challenges in capturing high-frequency, repetitive transactions and long-term trading patterns. In contrast, sequence models that directly handle transaction data largely avoid these issues. In response, the pre-trained Transformer-based account encoder, BERT4ETH \cite{bert4eth}, was proposed. It leverages the Transformer architecture, commonly used in language models, to handle temporally ordered transaction events and employs a masked language model for pre-training, where transaction addresses are randomly masked. The model is then fine-tuned with an MLP for account classification. ZipZap \cite{zipzap} builds on BERT4ETH by introducing frequency-aware compression techniques, aimed at reducing the cost of using BERT4ETH while maintaining similar performance. 

However, despite borrowing from BERT's architecture and pre-training methods, BERT4ETH processes context-free data structures, which hampers its ability to learn semantic information. Additionally, BERT4ETH masks transaction addresses rather than financial records directly related to transaction patterns, leading the model to focus on account relationships and overlook transaction details.

\section{Method}
In this section, we provide a explanation of our proposed LMAE4Eth framework, detailing the construction of the  pre-trained transaction language model, the masked account graph auto-encoder and the cross-attention fusion 
network. For clarity, our explanation is structured into three subsections:
\begin{itemize}[leftmargin=*]
\item Transaction-token Contrastive Language Model: This subsection explains how we build a language model for transactions to capture semantic information and create semantically distinguishable account representations. The core components include the semantic representations of transaction sequences and a token-aware contrastive learning pretraining task.
\item Masked Account Graph Auto-Encoder: This subsection discusses how we construct a self-supervised graph representation learning model to overcome the limitations of previous account transaction graph models. Key components include node initialization based on expert-engineered features and a graph mask autoencoder with layer-neighbor sampling.
\item Cross-Attention Fusion Network: This subsection explains our method for fusing pre-trained semantic embeddings and self-supervised graph embeddings enhanced with expert knowledge into a unified representation.
\end{itemize}

\subsection{Transaction-token Contrastive Language Model}
Our transaction-token contrastive language model (TxCLM) adopts a BERT-consistent Transformer architecture, constructed through two pivotal components:1) Linguistically Interpretable Representation: We transform context-independent digital data into linguistically interpretable formats. 2) Self-Supervised Pre-Training: We pretrain the model using token-aware contrastive learning and a masked language model. Note: In this section, we focus on the semantic indicative representation of transaction records and token-aware contrastive learning, and do not revisit the BERT Transformer architecture and masked language model. 

\subsubsection{\textbf{Linguistically Interpretable Representation of Transactions}}

Previous detection models faced challenges in collecting sufficient labeled data for Ethereum fraud detection, as they rely on multi-round supervised optimization to classify accounts into predefined categories (currently, only about 7,000 publicly disclosed phishing addresses are available). The pre-trained Transformer-based account encoder addresses this challenge by using unsupervised pretraining to encode transaction sequences, and overcomes the limitations of earlier graph-based account learning methods that struggled to capture high-frequency transactions and long-term dependencies. However, existing pre-trained Transformer account encoders process context-independent numerical inputs, making it difficult to learn the semantics of transactions. Additionally, the multi-variable nature of such data can cause the masked language model to fail. To address this, we propose converting transaction records into language-interpretable representations, enabling the pre-trained Transformer to leverage the powerful information encoding capabilities of language models to uncover insights that are not easily revealed by raw numerical data.

Let $\mathcal{T} = \{ t_1, t_2, \ldots, t_N \}$ denote the set of $N$ transactions associated with a single account. Each transaction $t_i$ is characterized by a tuple:
\begin{equation}
    t_i = ( v_i, d_i, \tau_i )
\end{equation}
where:
\begin{itemize}
    \item $v_i$ represents the transaction amount.
    \item $d_i \in \{ -1, 1 \}$ denotes the transaction direction, with $-1$ indicating an inflow and $1$ indicating an outflow.
    \item $\tau_i \in \mathbb{T}$ is the timestamp of the transaction, where $\mathbb{T}$ represents the set of all possible timestamps.
\end{itemize}

To transform these numerical attributes into a format compatible with a language model, we convert each transaction into a set of linguistic tokens by prepending descriptive text indicators to the attributes:
\begin{equation}
    \mathcal{L}( t_i ) = \{ \text{amount: } v_i, \text{ direction: } d_i, \text{ timestamp: } \tau_i \}
\end{equation}
We exclude Ethereum addresses from the linguistically interpretable representation. Ethereum addresses are hexadecimal strings of 40 characters prefixed with ``0x''. Incorporating such lengthy strings poses challenges for standard tokenizers, which are typically unable to segment these addresses into meaningful subunits. Including such long addresses in the linguistically interpretable transaction representation significantly reduces the number of transactions that can be processed by the model and introduces noisy information due to improper tokenization. Instead, we delegate the representation of account interaction features involving addresses to the graph model, which is better suited for capturing relational and structural information.

The $N$ transactions of an account are sequentially organized into a series of transaction sentences $\mathcal{C}$:
\begin{equation}
    \mathcal{C} = \{ \mathcal{L}( t_1 ), \mathcal{L}( t_2 ), \ldots, \mathcal{L}( t_N ) \}
\end{equation}
This transformation enables the language model to process the transaction data as a structured text corpus, allowing it to learn semantic representations from the transactional context and relationships within the account's activity.

\subsubsection{\textbf{Token-aware Contrastive Learning}}

Transactions on Ethereum are highly structured and homogeneous, which causes the encoder to cluster the semantic features of tokens into narrow subspaces within the feature space when learning transaction sequences. This limits the prominence of semantic differences. However, for fraud detection tasks, certain labels (specific terms within the transaction records) often reveal potential semantic signals of abnormal activity. To address this issue, we introduce a token-aware contrastive learning pretraining task to capture subtle differences in transaction attributes between accounts, enabling the learning of more expressive account representations.

Specifically, We adopt an anchor-enhanced architecture, where both the anchor model \( A \) and the enhanced model \( E \) are initialized with the same pretrained BERT weights. During pretraining, the anchor model \( A \) remains frozen, while the enhanced model \( E \) is optimized based on a set of objectives. Notably, we remove the NSP objective from the original BERT pretraining tasks, as each account's transaction records are treated as a single sentence in our data, and there is no meaningful relationship between sentences. Instead, we introduce a token-aware contrastive learning objective \( \mathcal{L}_{Ta} \) to complement the MLM objective.

Given the input sequence $\mathcal{C} = \{ L( t_1 ), L( t_2 ), \ldots, L( t_N ) \}$ representing the transaction sentences of an account. We randomly mask tokens in $\mathcal{C}$ using the same strategy as BERT to produce the masked sequence $\tilde{\mathcal{C}}$. The enhanced model \( E \) takes $\tilde{\mathcal{C}}$ as input and produces contextual representations $\tilde{h} = [\tilde{h}_0, \tilde{h}_1, \ldots, \tilde{h}_n]$, where $\tilde{h}_i \in \mathbb{R}^{d_\text{LM}}$ is the embedding of the \( i \)-th token in $\tilde{\mathcal{C}}$. Simultaneously, the anchor model \( A \) processes the original sequence $\mathcal{C}$ to generate reference representations $h = [h_0, h_1, \ldots, h_n]$, where $h_i \in \mathbb{R}^{d_\text{LM}}$ is the embedding of the \( i \)-th token in $\mathcal{C}$.

The representations of masked tokens from the enhanced model are compared to their corresponding reference representations from the anchor model. The enhanced model is trained to make the masked token representations closer to the anchor model’s reference representations, while pushing them away from other token representations in the same sequence. The contrastive learning objective is defined as:
\begin{equation}
\mathcal{L}_{\text{Ta}} = -\sum_{i=1}^{n} \textup{mask}(\tilde{x}_i) \log \frac{\exp(\textup{sim}(\tilde{h}_i, h_i) / \tau)}{\sum_{j=1}^{n} \exp(\textup{sim}(\tilde{h}_i, h_j) / \tau)}
\end{equation}
where \( \textup{mask}(\tilde{x}_i) = 1 \) if \( \tilde{x}_i \) is a masked token and \( \textup{mask}(\tilde{x}_i) = 0 \) otherwise. The function \( \textup{sim}(\cdot, \cdot) \) measures the similarity between vectors, and \( \tau \) is a temperature parameter that controls the sensitivity of the loss to similarity.

The token-aware contrastive learning objective \( \mathcal{L}_{\text{Ta}} \) is paired with a standard masked language model (MLM) objective. The MLM objective is defined as:
\begin{equation}
\mathcal{L}_{\text{MLM}} = \mathbb{E}_{\mathcal{C}} \left[ - \sum_{i \in \mathcal{M}} \log P( x_i | \tilde{\mathcal{C}} ) \right]
\end{equation}
where $\mathcal{M}$ is the set of masked token indices, $\tilde{\mathcal{C}}$ is the masked sequence, $x_i$ is the \( i \)-th token in $\mathcal{C}$, and $P( x_i | \tilde{\mathcal{C}} )$ is the probability of predicting the original token $x_i$ given the masked context.

The overall learning objective combines the contrastive learning loss \( \mathcal{L}_{\text{Ta}} \) with the MLM loss \( \mathcal{L}_{\textup{MLM}} \):
\begin{equation}
\mathcal{L}_{\textup{TxCLM}} = \mathcal{L}_{\textup{MLM}} + \mathcal{L}_{\text{Ta}}
\end{equation}
where \( \mathcal{L}_{\textup{MLM}} \) captures the semantic relationships within transaction contexts, and \( \mathcal{L}_{\text{Ta}} \) ensures that the representations of masked tokens in the enhanced model are closer to the anchor representations while being distinct from other tokens in the feature space. Intuitively, \( \mathcal{L}_{\textup{MLM}} \) encourages the model to enhance its understanding of contextual information within transaction sequences, while \( \mathcal{L}_{\text{Ta}} \) promotes the generation of more discriminative embeddings for individual tokens in the sequence. This combined approach produces token-aware representations that are semantically diverse and aligned with latent fraud indicators in transaction data.
\subsection{Masked Account Graph Auto-Encoder}
\begin{table*}[htbp]
\small
\centering
\renewcommand{\arraystretch}{1.05}
\caption{The expert-engineered features used to initialize the nodes and their descriptions}
\setlength{\tabcolsep}{5mm}  
\begin{tabular}{ll}
\noalign{\hrule height 1.5pt}
\textbf{Feature} & \textbf{Description} \\ \hline
Node outdegree & Number of outgoing transactions from the account. \\ 
Node indegree & Number of incoming transactions to the account. \\ 
Max outgoing amount & Maximum transaction amount sent by the account. \\ 
Min outgoing amount & Minimum transaction amount sent by the account. \\ 
Max incoming amount & Maximum transaction amount received by the account. \\ 
Min incoming amount & Minimum transaction amount received by the account. \\ 
Average outgoing amount & Average transaction amount sent by the account. \\ 
Average oncoming amount & Average transaction amount received by the account. \\ 
Account balance &  The sum of all income minus the sum of outgoing \\ 

Account lifetime & Duration of time the account has been active. \\ 

Long-term incoming transfer frequency &The number of incoming transmissions to a node over a long-term window. \\

Short-term incoming transfer frequency &The number of incoming transmissions to a node over a short window.\\ 

Long-term outgoing transfer frequency &The number of outgoing transmissions from a node over a long-term window\\

Short-term outgoing transfer frequency &The number of outgoing transmissions from a node in a short window\\

Degree centrality & The number of connections a node has in the graph. \\ 
Indegree centrality & The number of incoming connections to a node. \\ 
Outdegree centrality & The number of outgoing connections from a node. \\ 
Betweenness centrality & Measures how often a node appears on the shortest paths between other nodes. \\ 
Closeness centrality & Measures the average shortest path from a node to all other nodes. \\ 
Eigenvector centrality & Measures the influence of a node in the graph based on its neighbors' importance. \\ 
Katz centrality & Measures the relative influence of a node by considering all paths from it. \\ 
Clustering coefficient & Measures the degree to which nodes in a graph tend to cluster together. \\ 
\noalign{\hrule height 1.5pt}
\end{tabular}
\renewcommand{\arraystretch}{1.2}
\label{tab:node_features}
\end{table*}
Graph-based models have demonstrated effectiveness in capturing intricate node relationships and the interdependencies within account transactions. However, current graph-based detection methods face technical limitations that reduce the quality of node embeddings and present scalability challenges. To address these issues and leverage the graph perspective, we propose a masked account graph autoencoder with layer-neighbor sampling (LABOR-MAGAE), which uses expert-engineered features to initialize node embeddings in the account graph.

\subsubsection{\textbf{Initialization of nodes with expert-engineered features}}

In the constructed account interaction graph \( G = (V, E) \), each node \( v \in V \) represents an Ethereum account, and each edge \( e = (u, v) \in E \) denotes a transactional relationship between two accounts \( u \) and \( v \). The weight \( w_{uv} \) of an edge \( e \) is proportional to the number of transactions between accounts \( u \) and \( v \). To apply GNNs on \( G \), it is essential to initialize node features \( \mathbf{x}_i \) for each \( v_{i} \in V \).

Existing graph-based Ethereum fraud detection methods initialize node features in three main ways:

\begin{itemize}[leftmargin=*]
\item  Using transaction data, such as the total transaction amount, total number of incoming transactions and outcoming transactions.
\item  Node embeddings generated by random walk methods, such as Node2Vec and Trans2Vec.
\item  Subgraph-based approaches, where local transaction subgraphs are extracted, and node classification is transformed into a graph classification problem.
\end{itemize}

Compared to the previous methods, we craft more detailed expert-engineered features by incorporating both empirical and statistical knowledge. For instance, fraudulent accounts often exhibit abnormal behavior in features such as account lifecycle, transaction frequency, and centrality measures compared to normal accounts. These features can be summarized into three categories: transaction statistics, transaction transfer behavior, and node importance metrics. As shown in Table~\ref{tab:node_features}, our expert-engineered features include transaction statistics like in-degree, out-degree, maximum and minimum expenditure, maximum and minimum income, average expenditure and income, and account lifecycle. To quantify node importance, we calculate centrality measures such as degree centrality, in-degree centrality, out-degree centrality, betweenness centrality, closeness centrality, eigenvector centrality, Katz centrality, and clustering coefficient. Additionally, we analyze transaction transfer behavior by calculating the short-term and long-term frequencies of incoming and outgoing transactions. These features together provide a comprehensive view of token transfer structures and account interactions within the Ethereum network, aiding in the detection of anomalies or fraudulent behavior.

Finally, each node \( v \) is represented by \( \mathbf{x}_v \in \mathbb{R}^{d_\text{Node}}\), capturing transactional statistics, importance measures, and behavioral features. Edges \( e \in E \) are assigned weights \( w_{uv} \) proportional to the number of transactions between nodes.

\subsubsection{\textbf{LABOR-MAGAE}}

Graph embedding and Graph Neural Networks (GNNs) are widely used in Ethereum transaction networks. However, shallow embedding methods like graph embeddings are limited in capturing complex node representations. Most GNNs, based on graph convolution or message-passing mechanisms, suffer from neighborhood explosion or over-smoothing, which degrades the quality of node embeddings. Recent research has explored graph classification algorithms or traditional Graph Autoencoders (GAE)\cite{VGAE}, but these methods predominantly focus on structural reconstruction. While this approach ensures reliable link prediction and node clustering performance \cite{meng2019co-embedding}, it yields suboptimal results for node classification tasks \cite{yang2024vgae_for_Semi-Supervised_Classification,graphmae,li2024revisiting_and_Benchmarking_GAE,tan2023s2gae}. Furthermore, due to the large number of Ethereum transaction nodes, these methods are typically limited to smaller graphs and struggle to scale for large-scale training.

To address these challenges, we propose the LABOR-MAGAE, which combines Layer-Neighbor Sampling (LABOR) \cite{labor} with Masked Graph Autoencoder \cite{graphmae}. This approach focuses on node feature reconstruction rather than aggregating neighborhood information or reconstructing the graph structure, thus avoiding issues like neighborhood explosion and over-smoothing. Additionally, hierarchical node sampling enables efficient large-scale training. A detailed overview of the LABOR-MAGAE method is as follows:

\noindent\textbf{Input Data:}
The account interaction graph \( G = (V, E) \) constructed from Ethereum transaction records, Each node \( v_{i} \in V \) represents an account, and its initial feature vector \( \mathbf{x}_i \in \mathbb{R}^{d_\text{Node}} \) is derived from transactional statistics, capturing historical behaviors of accounts. Each edge \( (u, v) \in E \) represents a transaction from account \( u \) to account \( v \), and the edge weight \( w_{uv} \) is the normalized transaction count.

\noindent\textbf{Graph encoding with LABOR Sampling:}
To begin with, the graph \( G = (V, E) \) is partitioned into \( B \) mini-batches, where each mini-batch \( V_b \subseteq V \) contains a subset of nodes. These mini-batches satisfy the condition that the union of all mini-batches covers the entire set of nodes:
\begin{equation}
V = \bigcup_{b=1}^B V_b, \quad V_i \cap V_j = \emptyset \text{ for } i \neq j
\end{equation}
For each mini-batch \( V_b \), the LABOR sampling mechanism is applied to sample neighbors for each node \( v \in V_b \), where the set of sampled neighbors for node \( v \) is defined as:
\begin{equation}
N_{\text{LABOR}}(v) = \{u \in N(v) \mid r_u \leq c_v \pi_u\}
\end{equation}
Here, \( N(v) \) represents the full set of neighbors for node \( v \), \( \pi_u \) is the sampling probability for neighbor \( u \), \( c_v \) is the normalization factor, and \( r_u \sim U(0, 1) \) is a random variable. Intuitively, for each neighbor \( u \in N(v) \), the inclusion of \( u \) in the sample set is determined with probability \( c_v \pi_u \).

The adjacency matrix for the mini-batch is formed by aggregating the sampled neighbors for each node:
\begin{equation}
A_{\text{LABOR}, b} = \bigcup_{v \in V_b} N_{\text{LABOR}}(v)
\end{equation}
This ensures that all nodes \( V \) are covered in one epoch.

Next, a subset of nodes \( \tilde{V}_b \subset V_b \) is randomly selected, and their feature vectors are replaced with a learnable mask token [MASK], \( \mathbf{x}_{\text{mask}} \in \mathbb{R}^{d_h}\), the node feature $\mathbf{x}_i$ for $v_i\in\mathbf{V}$ in the masked feature matrix $\tilde{\mathbf{X}}_i$ can be defined as:
\begin{equation}
\tilde{\mathbf{X}}_i = 
\begin{cases} 
\mathbf{x}_i, & \text{if } i \notin \tilde{V}_b \\
\mathbf{x}_{\text{mask}}, & \text{if } i \in \tilde{V}_b
\end{cases}
\end{equation}
The encoder \( f_E \) takes the sampled adjacency matrix \( A_{\text{LABOR}, b} \) and the masked feature matrix \( \tilde{\mathbf{X}}_b \) as input to compute hidden representations:
\begin{equation}
H_b = f_E(A_{\text{LABOR}, b}, \tilde{\mathbf{X}}_b) \in \mathbb{R}^{|V_b| \times d_h}
\end{equation}
where \( d_h \) is the latent feature dimension. The representations for all mini-batches are then aggregated to form the complete hidden representation matrix:
\begin{equation}
H = \bigcup_{b=1}^B H_b \in \mathbb{R}^{|V| \times d_h}
\end{equation}
\noindent\textbf{Graph decoding with LABOR Sampling:}
During the decoding stage, the hidden representation of the masked nodes is again replaced with a decoding-specific token \( H_{\text{DMASK}} \), as described in the following equation:
\begin{equation}
\tilde{H}_i = 
\begin{cases} 
H_i, & \text{if } i \notin \tilde{V}_b \\
H_{\text{DMASK}}, & \text{if } i \in \tilde{V}_b
\end{cases}
\label{eq:re_masking}
\end{equation}
The decoder \( f_D \) then takes the adjacency matrix \( A_{\text{LABOR}, b} \) and the re-masked features \( \tilde{H}_b \) as input to reconstruct the original features. The output of the decoder is given by:
\begin{equation}
Z_b = f_D(A_{\text{LABOR}, b}, \tilde{H}_b) \in \mathbb{R}^{|V_b| \times {d_\text{Node}}}
\label{eq:decoding}
\end{equation}
Finally, the reconstructed features for all mini-batches are aggregated as follows:
\begin{equation}
Z = \bigcup_{b=1}^B Z_b \in \mathbb{R}^{|V| \times {d_\text{Node}}}
\label{eq:aggregated_decoding}
\end{equation}
This aggregation ensures that the model captures and combines the information from all batches in a unified representation.




\noindent\textbf{Loss Function:}
The reconstruction loss is computed for the masked nodes \( \tilde{V} = \bigcup_{b=1}^B \tilde{V}_b \) using the Scaled Cosine Error (SCE):
\begin{equation}
L_{\text{SCE}} = \frac{1}{|\tilde{V}|} \sum_{i \in \tilde{V}} \left( 1 - \frac{\mathbf{x}_i^\top \mathbf{z}_i}{\|\mathbf{x}_i\| \|\mathbf{z}_i\|} \right)^\gamma
\label{eq:loss_function}
\end{equation}
where: \( \mathbf{x}_i \) and \( \mathbf{z}_i \) are the original and reconstructed features for node \( i \), \( \|\cdot\| \) is the \( \ell_2 \)-norm, and \( \gamma \geq 1 \) is a scaling factor.

\noindent\textbf{Output:}
After node reconstruction pretraining, during the inference phase, the decoder \( f_D \) is discarded and the encoder \( f_E \) is applied to the Ethereum interaction graph \( G = (V, E) \) without any masking, generating interaction-based embeddings \( \hat{\mathbf{x}}_i \in \mathbb{R}^{d_h}\) of each node $v_i $  that capture the learned Ethereum network structure and key transaction patterns.

\section{Cross-Attention Fusion Network}

Our TxCLM generates semantic embeddings for accounts, $\mathbf{s} \in \mathbb{R}^{N \times d_{LM}}$, while LABOR-MAGAE produces account node embeddings, $\hat{\mathbf{x}} \in \mathbb{R}^{d_h}$, from the account transaction graph. These embeddings capture rich transaction semantics, expert-engineered account features and inter-account transaction dependencies. To fully leverage the complementary advantages of these different views, The network enables dynamic interactions between different views by  learnable fusion tokens and a cross-attention mechanism, as opposed to the linear layer fusion methods.
\begin{algorithm}
\caption{The process of Cross-Attention Fusion Network (CAFN)}
\label{alg:CAFN}

\KwIn{\,\,Semantic embeddings - \( \mathbf{S} \in \mathbb{R}^{N \times d_s} \) \\
Interaction embeddings - \( \hat{\mathbf{X}} \in \mathbb{R}^{|\mathcal{A}| \times d_g} \) \\
Labels - \( \mathcal{Y} \), Learning rate - \( \eta \) \\
Batch size - \( B \), Number of epochs - \( E \)}
\KwOut{Trained CAFN parameters \( \Theta \)}

\SetKwFunction{FMain}{CAFN}
\SetKwProg{Fn}{Function}{:}{}
\Fn{\FMain{}}{
    \(\triangleright\) Initialize CAFN parameters \\
    \( \Theta \gets \{\mathbf{A}^s, \mathbf{A}^f, \mathbf{W}_Q, \mathbf{W}_K, \mathbf{W}_V, \mathbf{W}_s, \mathbf{W}_g, \mathbf{b}, \Theta_{\text{MLP}}\} \)
    
    \For{\( \text{epoch} = 1 \) \KwTo \( E \)}{
        \For{\( \{(a_i, y_i)\}_{i=1}^B \in \text{MiniBatches}(\mathcal{A}, \mathcal{Y}) \)}{
            \(\triangleright\) Semantic aggregation for each account \( a_i \) \\
            \( \mathbf{Z}^s_{a_i} \gets \mathrm{CrossAttention}(\mathbf{A}^s, \mathbf{S}_{a_i}; \Theta) \)
            
            \(\triangleright\) Cross-perspective fusion \\
            \( \mathbf{Z}^{sg}_{a_i} \gets \sigma(\mathbf{W}_s \mathbf{Z}^s_{a_i} + \mathbf{W}_g \hat{\mathbf{x}}_{a_i} + \mathbf{b}) \)
            
            \(\triangleright\) Final fusion \\
            \( \mathbf{F}_{a_i} \gets \mathrm{CrossAttention}(\mathbf{A}^f, \mathbf{Z}^{sg}_{a_i}; \Theta) \)
            
            \(\triangleright\) Classification and loss computation \\
            \( \hat{y}_{a_i} \gets \mathrm{MLP}(\mathrm{Pool}(\mathbf{F}_{a_i}); \Theta_{\text{MLP}}) \) \\
            \( \mathcal{L}_{a_i} \gets \mathrm{CrossEntropyLoss}(\hat{y}_{a_i}, y_i) \)
        }
        \(\triangleright\) Update parameters \\
        \( \Theta \gets \Theta - \eta \nabla_{\Theta} \sum_{i=1}^B \mathcal{L}_{a_i} \)
    }
}
\KwRet Trained parameters \( \Theta \)
\end{algorithm}
Given semantic embeddings \( \mathbf{s} \in \mathbb{R}^{N \times d_{\text{LM}}} \) and interaction embeddings \( \hat{\mathbf{x}} \in \mathbb{R}^{d_h} \) of an account, both are inconsistent in terms of embedding dimension and sequence length. Specifically, semantic embeddings have a sequence length \( N \) (the number of tokens in the transaction text) and embedding dimension \( d_{\text{LM}} \), while interaction embeddings have a fixed embedding dimension \( d_h \) but do not have a sequence structure. In order to unify the shape of features from these two perspectives while ensuring embedding quality and computational efficiency, we introduce factorized feature compression based on aggregate tokens \cite{tokenfusion1,tokenfusion2} and cross-attention mechanism. For convenience, we will use \( d_s \) to denote \( d_{\text{LM}} \) and \( d_g \) to denote \( d_h \) in the following discussions.

Specify, we introduce a set of learnable semantic aggregate tokens \( \mathbf{A}^s \in \mathbb{R}^{k_s \times d_s} \), where \( k_s \) is the number of aggregate tokens used for the semantic features and \( d_s \) is the semantic embedding dimension. These tokens interact with the semantic embeddings through cross-attention \cite{attention}to generate compressed representations:
\begin{equation}
    \mathbf{Z}^s = \text{CrossAttention}(\mathbf{A}^s, \mathbf{S}) = \text{softmax}\left(\frac{\mathbf{Q}_s \mathbf{K}_s^\top}{\sqrt{d_s}}\right) \mathbf{V}_s
\end{equation}
\[
    \mathbf{Q}_s = \mathbf{A}^s \mathbf{W}_Q^s, \quad \mathbf{K}_s = \mathbf{S} \mathbf{W}_K^s, \quad \mathbf{V}_s = \mathbf{S} \mathbf{W}_V^s
\]
where \( \mathbf{Q}_s \), \( \mathbf{K}_s \), and \( \mathbf{V}_s \) are the query, key, and value matrices derived from the semantic embeddings \( \mathbf{S} \). The output of this cross-attention operation, \( \mathbf{Z}^s \in \mathbb{R}^{k_s \times d_s} \), is a compressed representation of the semantic features, where the number of tokens \( k_s \) is much smaller than the original sequence length \( N \), thus reducing the computational complexity.

Next, we compute the cross-perspective interaction representations by fusing the aggregated semantic features \( \mathbf{z}_i^s \in \mathbb{R}^{d_s} \) with the interaction embeddings \( \hat{\mathbf{x}}_i \in \mathbb{R}^{d_h} \) for each corresponding account. To fuse these features, we apply a linear layer:
\begin{equation}
    \mathbf{Z}^{sg}_i = \sigma\left(\mathbf{W}_s \mathbf{z}_i^s + \mathbf{W}_g \hat{\mathbf{x}}_i + \mathbf{b}\right)
\end{equation}
where \( \mathbf{W}_s \in \mathbb{R}^{d_f \times d_s} \) and \( \mathbf{W}_g \in \mathbb{R}^{d_f \times d_h} \) are the projection matrices for the semantic and interaction features, respectively, \( \mathbf{b} \in \mathbb{R}^{d_f} \) is a bias term, and \( \sigma(\cdot) \) is a non-linear activation function. This operation generates the cross-perspective interaction representations \( \mathbf{Z}^{sg}_i \) for each pair of semantic and interaction features.
\begin{equation}
    \mathbf{Z}^{sg} = [\mathbf{Z}^{sg}_1, \mathbf{Z}^{sg}_2, \dots, \mathbf{Z}^{sg}_{k_s}] \in \mathbb{R}^{k_s \times d_f}
\end{equation}
where \( d_f \) is the dimensionality of the fused latent space, capturing multi-level interactions between the semantic and interaction features.

Finally, to generate the final fused account representation, we introduce learnable fusion tokens \( \mathbf{A}^f \in \mathbb{R}^{k_f \times d_f} \), which interact with the cross-perspective representations \( \mathbf{Z}^{sg} \) through another cross-attention mechanism:
\begin{equation}
    \mathbf{F} = \text{CrossAttention}(\mathbf{A}^f, \mathbf{Z}^{sg}) = \text{softmax}\left(\frac{\mathbf{Q}_f \mathbf{K}_f^\top}{\sqrt{d_f}}\right) \mathbf{V}_f
\end{equation}
\[
    \mathbf{Q}_f = \mathbf{A}^f \mathbf{W}_Q^f, \quad \mathbf{K}_f = \mathbf{Z}^{sg} \mathbf{W}_K^f, \quad \mathbf{V}_f = \mathbf{Z}^{sg} \mathbf{W}_V^f
\]
The final fused representation \( \mathbf{F} \in \mathbb{R}^{d_f} \) captures multi-level interactions between semantic and interaction embeddings. During fine-tuning for downstream tasks, \( \mathbf{F} \) is passed through a multi-layer perceptron (MLP) to obtain the classification results. The model is optimized using the cross-entropy loss function:
\begin{equation}
    L = -\sum_{c=1}^{ii} y_{ic} \log\left(\text{Softmax}\left(\text{MLP}(F_i  )\right)\right)_{ic}
\end{equation}
where \( M \) is the number of classes, \( y_{ic} \) represents the ground truth for class \( c \) of instance \( i \). The loss is used to optimize the parameters of the aggregation tokens \( \mathbf{A}^s \), fusion tokens \( \mathbf{A}^f \), and other parameters in the fusion module, such as \( \mathbf{W}_s \) and \( \mathbf{W}_g \), progressively improving the model's classification performance.

\section{Dataset Review }

\begin{table}
    \caption{Summary of datasets.}
    \renewcommand{\arraystretch}{1.2}
    \setlength{\tabcolsep}{2pt} 
    \centering
    \normalsize
    \begin{tabular}{ccccc}  
        \noalign{\hrule height 1.5pt}
        \textbf{Dataset} & \textbf{Nodes} & \textbf{Trans} & \textbf{Avg Degree} & \textbf{Phisher}\\ \hline
        MulDiGraph & 2,973,489 & 13,551,303 & 4.5574 & 1,165\\ 
        B4E & 597,258 & 11,678,901 & 19.5542 & 3,220\\ 
        SPN & 496,740 & 831,082 & 1.6730 & 5,619 \\ 
        \noalign{\hrule height 1.5pt}
    \renewcommand{\arraystretch}{1}
    \end{tabular}
    \vspace{-15pt}
    \label{tab:dataset}
\end{table}

As shown in Table \ref{tab:dataset}, we utiliz three datasets: MulDiGraph, B4E, and SPN. These datasets differ in terms of their collection methods, time spans, and organizational structures, allowing for the evaluation of the model's performance across various Ethereum transaction views.      

\begin{itemize}[leftmargin=*]
\item \textbf{MulDiGraph}\cite{xblock_1} A graph-structured dataset. This dataset is publicly available on the XBlock\cite{xblock} platform and is a widely used dataset that was released in December 2020. It includes a large Ethereum transaction network obtained by performing a two-hop Breadth-First Search (BFS) from known phishing nodes. The dataset contains 2,973,489 nodes, 13,551,303 edges, and 1,165 phishing nodes.

\item \textbf{B4E}\cite{bert4eth} A dataset of document structure. This dataset was collected via an Ethereum node using Geth \cite{bert4eth}. It covers transactions from January 1, 2017, to May 1, 2022, including 3,220 phishing accounts and 594,038 normal accounts. The dataset contains 328,261 transactions involving phishing accounts and 1,350,640 involving normal accounts. We constructed the corresponding transaction graph based on the transaction documents of B4E dataset during data preprocessing.

\item \textbf{SPN}\cite{spn} A graph-structured dataset. The SPN dataset constructed by first identifying known phishing nodes and performing a two-hop BFS to gather information on their neighboring nodes. The dataset contains trading information up until June 7, 2024, and includes 5,619 phishing accounts and 491,121 normal accounts, with a total of 831,082 transaction edges. SPN offers a current view of the Ethereum trading environment, focusing on recent phishing activities and the dynamics of the network.
\end{itemize}

\section{Experiment Section}
In this section, we evaluate the proposed method against 15 baselines. The experimental setup is as follows:

\begin{itemize}[leftmargin=*]
    \item Comparison of model performance using four metrics: precision, recall, F1-score, and balanced accuracy.
    \item Ablation experiments for each task module ,including token-aware contrastive learning (TxCLM) component, masked
account graph autoencoder (MAGAE) component, and cross-attention fusion network (CAFN) component, as well as an analysis of the complementary nature of features across modules.
    \item Comparison of the TxCLM and MAGAE components with other graph neural network models and language models, examining performance differences under various model architectures.
    \item Experiments with different-sized pretraining datasets to assess the model's generalization ability and dependence on data scale.
    \item Sensitivity analysis of key hyperparameters within the model.
\end{itemize}

comparative analysis encompasses 21 baseline models across three primary methodological domains. Graph embedding methods include DeepWalk~\cite{deepwalk}, Role2Vec~\cite{role2vec}, and Trans2Vec~\cite{trans2vec}. Transformer-based pre-trained approaches feature BERT4ETH~\cite{bert4eth}, ZipZap~\cite{zipzap},T5 \cite{raffel2020t5} and Longformer\cite{beltagy2020longformer}. Graph-based techniques comprise GCN~\cite{gcn}, GraphSAGE~\cite{gsage}, GRAND(N)\cite{chamberlain2021grand_Graph_neural_diffusion}, GRAND\cite{feng2020grand}, DiffPool~\cite{diffpool}, U2GNN~\cite{U2GNN}, IM\_Graph2Vec~\cite{im_graph2vec}, Graph2Vec~\cite{graph2vec}, TSGN~\cite{tsgn}, GrabPhisher~\cite{grabphisher}, CATALOG \cite{ghosh2025catalog}, VGAE~\cite{VGAE}, and GATE~\cite{gate}.  The graph-based category particularly spans a diverse range of methodological strategies, incorporating neighborhood aggregation, graph classification, structural reconstruction, and graph-centric information integration. We adopt a random node-based splitting strategy on the transaction graph, dividing the accounts into training, validation, and test sets with a ratio of 7:1:2.

\subsection{Comparison with Baselines}

\begin{table*}[htbp]
\caption{The performances with our method and baseline methods on three datasets, and BAcc is Balanced Accuracy.}
\renewcommand{\arraystretch}{1}
\setlength{\tabcolsep}{3pt} 
\normalsize 
\centering
\begin{tabular}{c|cccc|cccc|cccc}
\noalign{\hrule height 1.5pt} 
\multirow{2}[0]{*}{Method} & \multicolumn{4}{c|}{MulDiGraph} & \multicolumn{4}{c|}{B4E} & \multicolumn{4}{c}{SPN} \\ \cline{2-13}
&\textbf{Precision} & \textbf{Recall} & \textbf{F1} & \textbf{BAcc} & \textbf{Precision} & \textbf{ Recall} & \textbf{F1} & \textbf{BAcc} & \textbf{Precision} & \textbf{Recall} & \textbf{F1} & \textbf{BAcc} \\ \hline
DeepWalk & 0.5821  & 0.5867  & 0.5844  & 0.6880  & 0.6358  & 0.6495  & 0.6426  & 0.7317  & 0.5213  & 0.5119  & 0.5166  & 0.6384  \\
Role2Vec & 0.4688  & 0.6976  & 0.5608  & 0.6511  & 0.5748  & 0.7958  & 0.6673  & 0.7507  & 0.4521  & 0.7059  & 0.5512  & 0.6391  \\
Trans2Vec & 0.7114  & 0.6944  & 0.7029  & 0.7768  & 0.2634  & 0.7043  & 0.3842  & 0.3598  & 0.3928  & 0.7381  & 0.5134  & 0.5838  \\
GCN   & 0.2960  & 0.7513  & 0.4247  & 0.4289  & 0.5515  & 0.7508  & 0.6359  & 0.7228  & 0.5046  & 0.4973  & 0.5009  & 0.6266  \\
GAT   & 0.2689  & 0.7917  & 0.4014  & 0.3577  & 0.4729  & \underline{\textbf{0.8348}}  & 0.6038  & 0.6848  & 0.5083  & 0.7720  & 0.6130  & 0.6993  \\
GSAGE & 0.3571  & 0.3299  & 0.3430  & 0.5164  & 0.4589  & 0.5826  & 0.5134  & 0.6196  & 0.4557  & 0.5817  & 0.5110  & 0.6172  \\
DiffPool & 0.6475  & 0.5767  & 0.6101  & 0.7099  & 0.5767  & 0.5058  & 0.5389  & 0.6601  & 0.5592  & 0.5103  & 0.5336  & 0.6546  \\
U2GNN & 0.6218  & 0.6074  & 0.6145  & 0.7113  & 0.6236  & 0.5712  & 0.5963  & 0.6994  & 0.5766  & 0.5311  & 0.5529  & 0.6681  \\
Graph2Vec & 0.8293  & 0.4359  & 0.5714  & 0.6955  & 0.7714  & 0.4761  & 0.5888  & 0.7028  & \underline{\textbf{0.7951}}  & 0.4452  & 0.5708  & 0.6939  \\
TSGN  & \underline{\textbf{0.8544}}  & 0.5712  & 0.6847  & 0.7613  & 0.6233  & 0.8168  & 0.7071& 0.7850 & 0.7389  & 0.5128  & 0.6054  & 0.7111  \\
GrabPhisher & 0.7146  & \underline{\textbf{0.8472}}  & 0.7753 & 0.8390 & \underline{\textbf{0.8083}}  & 0.5931  & 0.6842  & 0.7614  & 0.6760  & \underline{\textbf{0.8059}}  & \underline{\textbf{0.7353}} & \underline{\textbf{0.8064}} \\
GAE   & 0.3728  & 0.5447  & 0.4426  & 0.5432  & 0.4239  & 0.5623  & 0.4834  & 0.5901  & 0.4077  & 0.3692  & 0.3875  & 0.5505  \\
GATE  & 0.3430  & 0.7138  & 0.4633  & 0.5151  & 0.4680  & 0.7191  & 0.5670  & 0.6552  & 0.6154  & 0.7376  & 0.6710  & 0.7536  \\
BERT4ETH & 0.4469  & 0.7344  & 0.5557  & 0.6400  & 0.7421  & 0.6125  & 0.6711  & 0.7530  & 0.7566  & 0.6713  & 0.7114  & 0.7817  \\
ZipZap & 0.4537  & 0.7298  & 0.5595  & 0.6452  & 0.7374  & 0.6132  & 0.6696  & 0.7520  & 0.7539  & 0.6682  & 0.7084 & 0.7796  \\ 
CATALOG & 0.8150  & 0.6351 & 0.7139  & 0.7815  & 0.7484  & 0.7045  & \underline{\textbf{0.7258}} &   \underline{\textbf{0.7930}} & 0.6890  & 0.7511  & 0.7187 & 0.7908  \\
    GraphMLP & 0.7876  & 0.5548  & 0.6510  & 0.7400  & 0.7647  & 0.2266  & 0.3496  & 0.5959  & 0.4580  & 0.6698  & 0.5440 & 0.6367  \\
    GRAND(N) & 0.7968  & 0.5902  & 0.6781  & 0.7575  & 0.7857  & 0.2833  & 0.4164  & 0.6223  & 0.7344  & 0.5032  & 0.5972 & 0.7061  \\
    GRAND & 0.8338  & 0.5377  & 0.6538  & 0.7421  & 0.6901  & 0.3199  & 0.4372  & 0.6240  & 0.7117  & 0.4798  & 0.5732 & 0.6913  \\
    T5   & 0.7890  & 0.7983  & \underline{\textbf{0.7936} } & \underline{\textbf{0.8458 }} & 0.7041  & 0.5820  & 0.6373  & 0.7299  & 0.7507  & 0.7087  & 0.7291 & 0.7955  \\
    Longformer & 0.7817  & 0.7359  & 0.7581  & 0.8166  & 0.7851  & 0.5989  & 0.6795  & 0.7585  & 0.7824  & 0.6767  & 0.7257 & 0.7913  \\ \hline
Ours  & \textbf{0.9024 } & \textbf{0.8889 } & \textbf{0.8960 } & \textbf{0.9204 } & \textbf{0.7903 } & \textbf{0.8397 } & \textbf{0.8143} & \textbf{0.8641} & \textbf{0.8123} & \textbf{0.7885} & \textbf{0.8002} & \textbf{0.8487}  \\
Improv. (\%) & 4.8   & 4.17  & 10.24 & 7.46  & -1.8  & 0.49  & 8.85  & 7.11  & 1.72  & -1.74 & 6.49  & 4.23 \\
\noalign{\hrule height 1.5pt} 
\end{tabular}%
\renewcommand{\arraystretch}{1}
\label{tab:baseline}%
\end{table*}%

As shown in Table \ref{tab:baseline}, our LMAE4Eth consistently outperforms all baseline methods across three different Ethereum datasets: MulDiGraph, B4E, and SPN.  Specifically, compared to the strongest baseline models, LMAE4Eth achieves a significant improvement of approximately 10.24\% and 7.46\% in F1-score and balanced accuracy on the MulDiGraph dataset, 8.85\% and 7.11\% on B4E, and 6.49\% and 4.23\% on SPN. These substantial performance gains demonstrate the effectiveness of our paradigm.

Furthermore, our proposed method demonstrates competitive robustness and generalizability across three distinct datasets. Our method maintain exceptional performance even under challenging conditions such as severe class imbalance and low fraudulent account representation. In contrast, baseline methods reveal substantial inconsistencies. Trans2Vec, for instance,  achieving a precision of 0.7114 on the MulDiGraph dataset, yet plummeting to a mere 0.2634 on the B4E dataset. Similarly, even best approaches like TSGN, on the MulDiGraph dataset, displays an overly conservative performance profile, with a precision of 0.8544 but a recall of only 0.5712.  Moreover, we observed a performance scaling effect of our approach directly correlated with data volume—a characteristic absent in other methods.

For graph embedding models, DeepWalk, Role2Vec, and Trans2Vec all leverage random walk algorithms to learn graph representations. However, DeepWalk and Role2Vec demonstrated a balanced performance and showed relatively robust results across three datasets of varying scales. In contrast, Trans2Vec improved random walk strategies with two sampling techniques—amount-Based and time-Based Biased Sampling—tailored for Ethereum transaction networks. While it performed better on the MulDiGraph dataset, its generalizability was limited. On other datasets (B4E and SPN), Trans2Vec showed a strong bias towards predicting fraudulent accounts, leading to high false positive rates. On B4E, this was especially evident, with F1 score and Balanced Accuracy dropping to 0.3842 and 0.3598, respectively. These results indicate that Trans2Vec’s sampling mechanisms are sensitive to network structure, limiting its adaptability to different topologies.

Among Graph Neural Network (GNN) approaches, GCN and GAT demonstrated the most pronounced bias towards classifying positive samples, specifically fraudulent accounts. On the MulDiGraph dataset,  GCN achieved a recall of 0.7513 with a precision of only 0.2960, while GAT reached a recall of 0.7917 with an even lower precision of 0.2689. These results unequivocally indicate a high false positive rate. GraphSAGE (GSAGE) exhibited even more pronounced vulnerabilities, particularly when confronted with severely imbalanced datasets. On the MulDiGraph dataset—characterized by the highest total node count but the lowest proportion of fraudulent nodes—GSAGE's recall and precision dropped to 0.3299 and 0.3571, respectively, demonstrating a strong tendency to classify majority class samples and effectively failing to identify minority class instances. Even the excellent graph representation learning models GraphMLP, GRAND(N) and GRAND have been challenged by considerable prediction bias. On the three public datasets, they all tend to classify negative samples and have low recall rates for positive samples. For example, on the MulDigraph dataset, their recall rates are only about 50\%-60\%. Overall, the three GNN methods are significantly impacted by network structure and label distribution differences. LMAE4Eth, however, mitigates these challenges with an unsupervised learning approach, demonstrating greater robustness across various network configurations.

Compared to graph classification methods like DiffPool and U2GNN, LMAE4Eth improves by over 20\%. In Ethereum transaction graphs, the transaction behavior of each account and its interactions with other accounts are crucial for determining whether the account is fraudulent. Graph classification methods overlook fine-grained node interactions, treating transaction graphs as whole and using global subgraph information. This limits their ability to capture complex internal behaviors, reducing their effectiveness in tasks like distinguishing normal and fraudulent accounts.

Among all information integration methods, CATALOG, TSGN and GrabPhisher perform best in the baseline models, benefiting from modeling both transaction graphs and time series. However, there remains a notable performance gap with LMAE4Eth, with differences exceeding 6\%-10\% on three datasets. We attribute this performance difference to LMAE4Eth's ability to extract account transaction semantics. While the precision of LMAE4Eth on the B4E dataset and its recall on the SPN dataset are slightly lower than the best-performing baselines, we observe that these baselines tend to optimize one metric at the expense of the other, resulting in an imbalanced prediction profile. In contrast, LMAE4Eth achieves a more favorable balance between precision and recall, yielding superior F1 across all datasets, which is essential for reliable fraud detection in practice.

For GAE and GATE models, the performance of GAE is suboptimal across all three datasets, while GATE shows significant improvements on the B4E and SPN datasets by incorporating attention mechanisms. However, as self-supervised graph learning algorithms, both GAE and GATE overly focus on graph structure and lack sufficient understanding of non-linear node interactions. In contrast, our MAGAE approach, which integrates node masking strategies and LABOR sampling, allows for more effective modeling of node interactions.

BERT4ETH and ZipZap are both pre-trained Transformer encoders that model relationships between accounts by capturing sequential patterns in account activities and predicting masked tokens. They perform excellently on the B4E and SPN datasets but experience a sharp decline in performance on MulDiGraph, underperforming compared to some graph-based methods. This suggests they are still influenced by dataset label quantity and class imbalance. As excellent sequence models, T5 and Longformer have stable performance on all datasets, especially T5 is the best baseline on MuDigraph. However, they still have a certain gap with our model and some information integration based methods. Compared to the best baseline models, we speculate that even advanced Transformer models require additional graph structure information to perform better.

\subsection{Ablation Study}

\begin{table}[t]
\small
\centering
\caption{Ablation study for LMAE4Eth.}
\setlength{\tabcolsep}{1.3mm}
\renewcommand{\arraystretch}{1}
\normalsize
\begin{threeparttable}
\begin{tabular}{c|cccc}
\noalign{\hrule height 1.5pt}
\textbf{Method}    & \textbf{Precision} &  \textbf{Recall}  & \textbf{F1}  & \textbf{BACC}  \\ \hline
w/o Language Model & 0.5724  & 0.7377  & 0.6446  & 0.7311  \\
w/o MAGAE & 0.7868  & 0.7703  & 0.7785  & 0.8330  \\
w/o Graph Model &  0.7652  & 0.7612  & 0.7632  & 0.8222  \\
w/o CAFN$^{\dag}$ &  0.7539  & 0.7712  & 0.7625  & 0.8227  \\
w/o CAFN$^{\ddagger}$ &  0.7384  & 0.7429  & 0.7406  & 0.8057  \\ \hdashline
w/o $\mathcal{L}_{\text{Ta}}$ & 0.7733 & 0.7962 & 0.7846  & 0.8397 \\
w/o statistics & 0.7568 & 0.8264 & 0.7902  & 0.8468 \\
w/o transfer behavior& 0.7842 & 0.8056 & 0.7947  & 0.8474 \\
w/o node importance& 0.7793 & 0.8427 & 0.8096  & 0.8617 \\
w/o all expert & 0.7701 & 0.8100  & 0.7894  & 0.8445 \\
\hline
Ours & \textbf{0.7903}  & \textbf{0.8397}  & \textbf{0.8143}  & \textbf{0.8641}  \\ 
\noalign{\hrule height 1.5pt}
\end{tabular}
\begin{tablenotes}
\footnotesize 
\item \ \ $\dag$ denotes the replacement of the CAFN with the direct addition.
\item $\ddagger$ denotes the replacement of the CAFN with the linear combination.
\end{tablenotes}
\end{threeparttable}
\renewcommand{\arraystretch}{1}
\label{tab:ablation_1}
\end{table}

As mentioned earlier, LMAE4Eth is built upon several important components, including Transaction Language Model, token-aware contrastive learning (TxCLM),  masked account graph autoencoder (MAGAE), and cross-attention fusion network (CAFN). To better understand the design principles of LMAE4Eth, we conduct a series of ablation experiments and analyses in this section.

Specifically, we apply the following ablations: 1) removing the token-aware contrastive learning module and replacing it with a basic BERT model (w/o TxCLM); 2) removing the semantic perspective in the CAFN module (w/o Language Model); 3) removing the node reconstruction strategy using the graph mask autoencoder and replacing it with a basic graph autoencoder (GAE); 4) removing the graph perspective  in the CAFN module (w/o Graph Model); 5) removing the CAFN (w/o CAFN), and instead using direct addition or linear combination of features from both perspectives.

Table \ref{tab:ablation_1} presents the results of these ablation experiments on the B4E dataset. It is evident that removing any component or strategy from the model leads to a performance drop, which demonstrates the effectiveness of our design choices. Among these, we find that the most significant performance degradation occurs when the semantic perspective transaction features are removed (w/o Language Model). Compared to the full model, F1 drops by 16.97\%, and Balanced Accuracy (BACC) decreases by 13.11\%. This indicates that the semantic perspective of Ethereum account transaction features plays a dominant role in the model's performance, and the language model proves to be highly effective in uncovering anomalous transaction behaviors.

Similarly, removing graph perspective (w/o Graph Model), which completely disregards the transactional relationships between accounts, also leads to a noticeable performance decline. On the B4E dataset, F1 drops by 5.11\%, and BACC decreases by 4.19\%. In fact, after repeated experiments and statistical analysis, we observed that removing graph perspective features causes the model to miss the fraudulent behaviors of approximately 110 accounts, which accounts for 3.4\% of the total phishing nodes in the B4E dataset. This underscores the indispensable nature of both perspectives in our model.

The CAFN, which incorporates the cross-attention mechanism and learnable fusion tokens, achieves an advantage of 5\% to 7\% in F1 compared to the simple addition and linear combination fusion methods. This also validates that our model effectively leverages the complementarity between semantic information and graph structural information, rather than just their individual advantages.

To further investigate the effect of token-aware contrastive learning, we introduce an ablation variant ``w/o~$\mathcal{L}_{\text{Ta}}$'', in which the language model is trained solely with the masked language modeling (MLM) objective. This variant yields an F1-score of 0.7846, representing a 2.97\% drop compared to the full model. We conjecture that this decline may indicate that the contrastive objective helps the model learn more discriminative token embeddings by encouraging it to distinguish subtle semantic variations that are potentially relevant for fraud detection.

Moreover, we ablate the expert-based features in three categories: transaction statistics (w/o statistics), transaction transfer behaviors (w/o transfer behavior), and node importance metrics (w/o node importance), resulting in F1-scores of 0.7902 (-2.41\%), 0.7947 (-1.96\%), and 0.8096 (-0.47\%), respectively. When all expert features are removed and node embeddings are randomly initialized (w/o all expert), the performance further drops to 0.7894 (-2.49\%). These results suggest that each expert-driven feature category contributes complementary discriminative signals, and their removal cumulatively impacts model performance.

\begin{table}[t]
\centering
\caption{Performance comparison of models after replacing MAGAE with other graph representation learning methods.}
\setlength{\tabcolsep}{2.8mm}
\renewcommand{\arraystretch}{1.05}
\normalsize
\centering
\begin{tabular}{c|cccc}
\noalign{\hrule height 1.5pt}
\textbf{Method}    & \textbf{Precision} &  \textbf{Recall}  & \textbf{F1}  & \textbf{BACC}  \\ \hline
w/ GCN & 0.7827  & 0.7782  & 0.7804  & 0.8351    \\
w/ GAT & 0.7667  & 0.8028  & 0.7843  & 0.8403    \\
w/ GSAGE & 0.7767  & 0.7889  & 0.7828  & 0.8377    \\
w/ GAE & 0.7866  & 0.7731  & 0.7798  & 0.8341    \\
w/ GATE & \underline{0.7919}  & \underline{0.8167}  & \underline{0.8041}  & \underline{0.8547}    \\ \hline
Ours & \textbf{0.7903}  & \textbf{0.8397}  & \textbf{0.8143}  & \textbf{0.8641}    \\ \noalign{\hrule height 1.5pt}
\end{tabular}
\renewcommand{\arraystretch}{1}
\label{tab:repalce_MAGAE}
\end{table}

\subsection{Architecture Advantage Analysis}

\begin{table}[htbp]
\centering
\caption{Performance comparison of models after replacing TxCLM with other language models.}
\setlength{\tabcolsep}{2mm}
\renewcommand{\arraystretch}{1.05}
\normalsize
\centering
\begin{tabular}{c|cccc}
\noalign{\hrule height 1.5pt}
\textbf{Method}    & \textbf{Precision} &  \textbf{Recall}  & \textbf{F1}  & \textbf{BACC}  \\ \hline
w/ BERT & \underline{0.8046}  & 0.7680  & 0.7859  & 0.8374  \\
w/ RoBERTa & 0.7792  & 0.7668  & 0.7730  & 0.8291    \\
w/ ALBERT & 0.7888  & \underline{0.7894}  & 0.7891  & 0.8419    \\
w/ ELECTRA & 0.7744  & 0.7692  & 0.7718  & 0.8286   \\
w/ VGCN-BERT & 0.8002  & 0.7859  & \underline{0.7930}  & \underline{0.8439}  \\ \hline
Ours & \textbf{0.7903}  & \textbf{0.8397}  & \textbf{0.8143}  & \textbf{0.8641}  \\ \noalign{\hrule height 1.5pt}
\end{tabular}
\renewcommand{\arraystretch}{1}
\label{tab:replace_TxCLM}
\end{table}

\begin{figure*}[htbp]
    \centering
    \begin{minipage}[b]{0.3\textwidth}
        \centering
        \includegraphics[width=\textwidth]{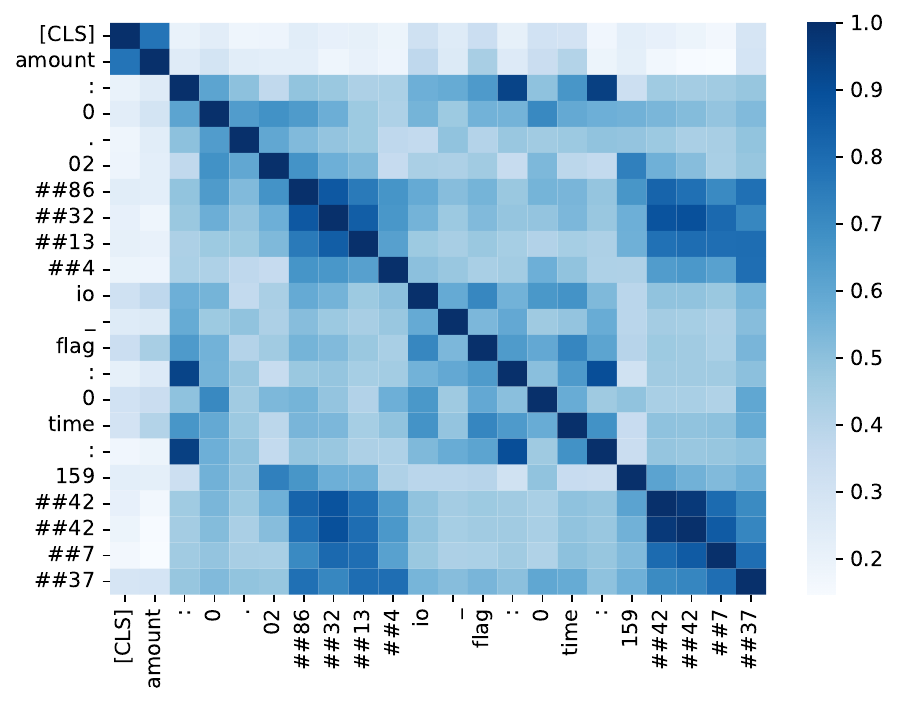}
    \end{minipage}
    \begin{minipage}[b]{0.3\textwidth}
        \centering
        \includegraphics[width=\textwidth]{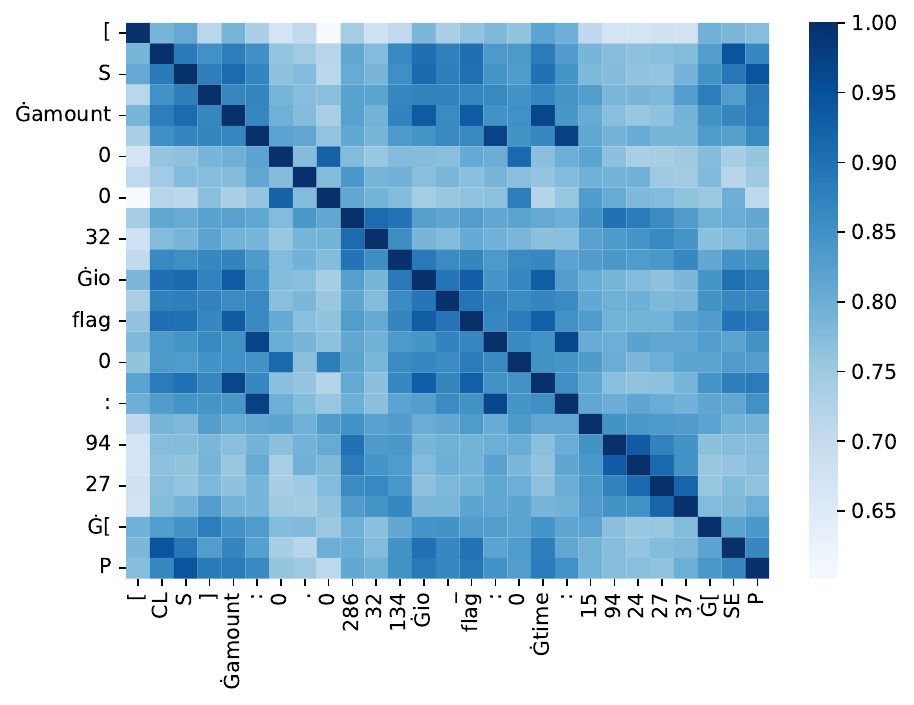}
    \end{minipage}
    \begin{minipage}[b]{0.3\textwidth}
        \centering
        \includegraphics[width=\textwidth]{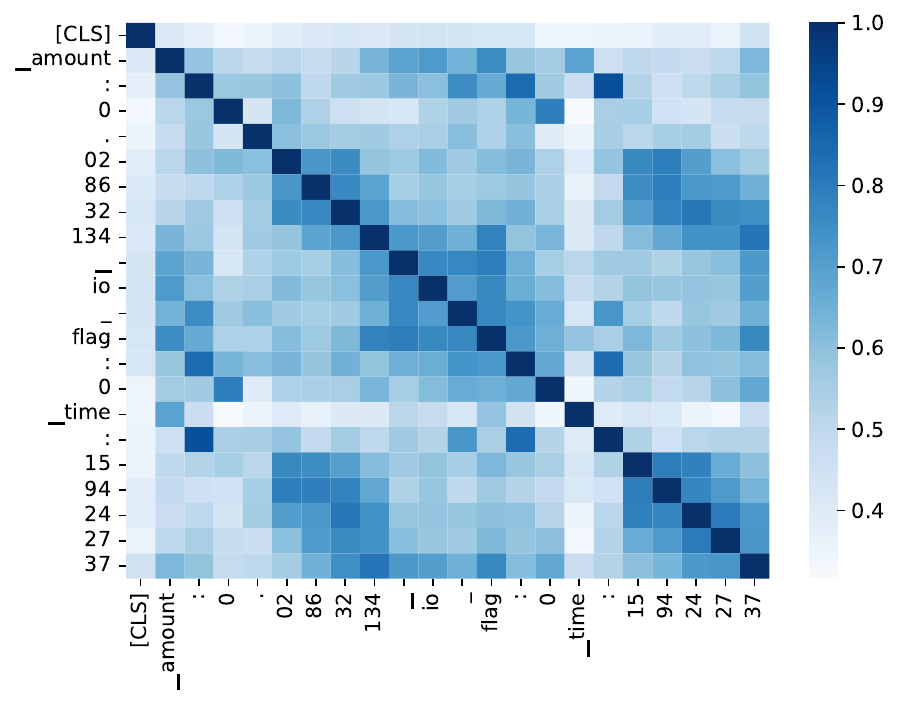}
    \end{minipage}

    \begin{minipage}[b]{0.3\textwidth}
        \centering
        \includegraphics[width=\textwidth]{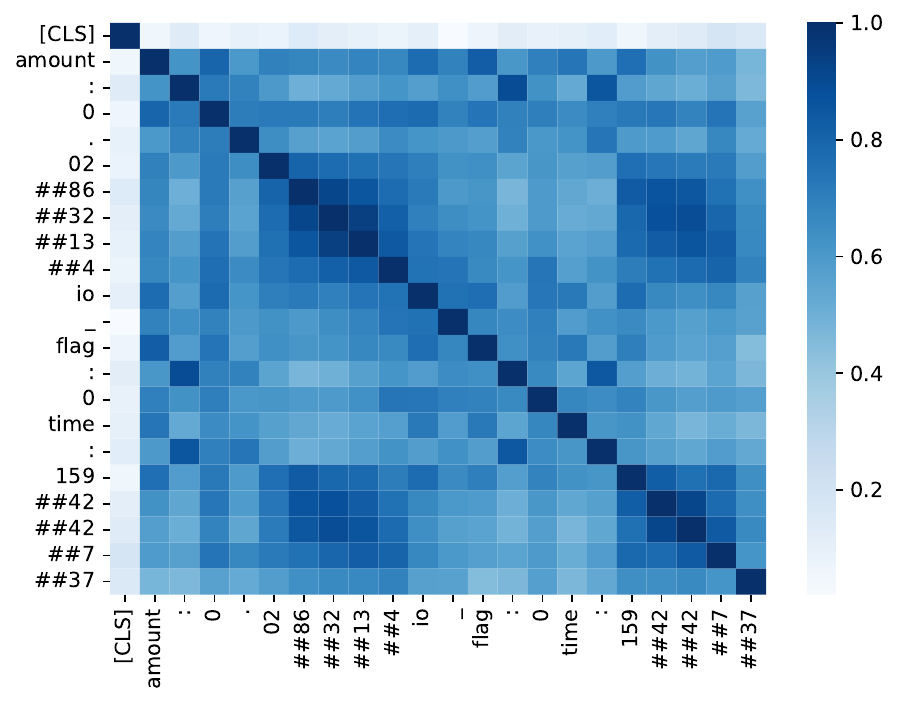}
    \end{minipage}
    \begin{minipage}[b]{0.3\textwidth}
        \centering
        \includegraphics[width=\textwidth]{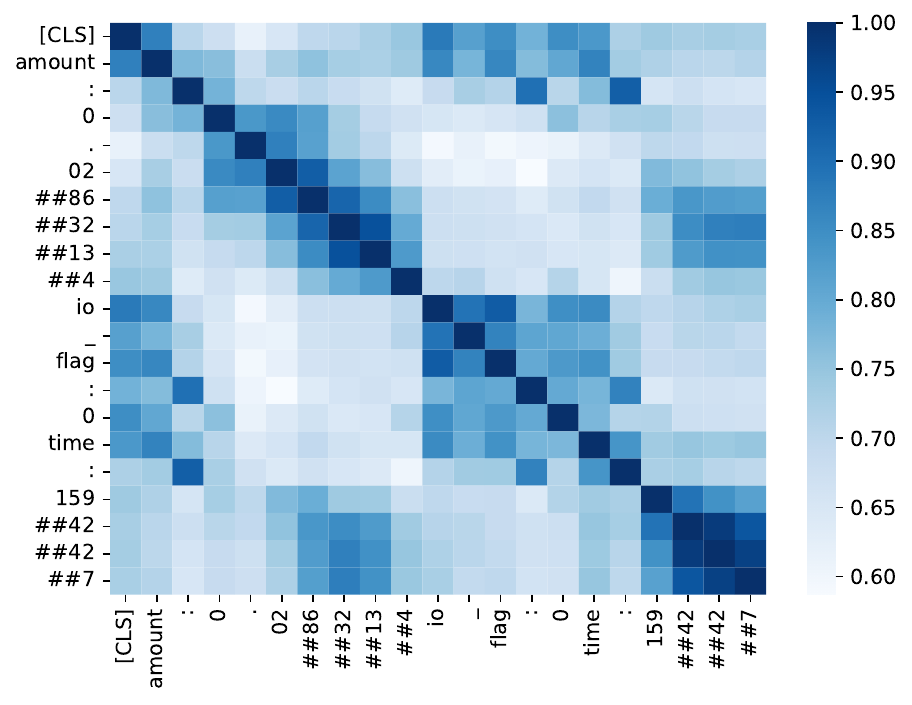}
    \end{minipage}
    \begin{minipage}[b]{0.3\textwidth}
        \centering
        \includegraphics[width=\textwidth]{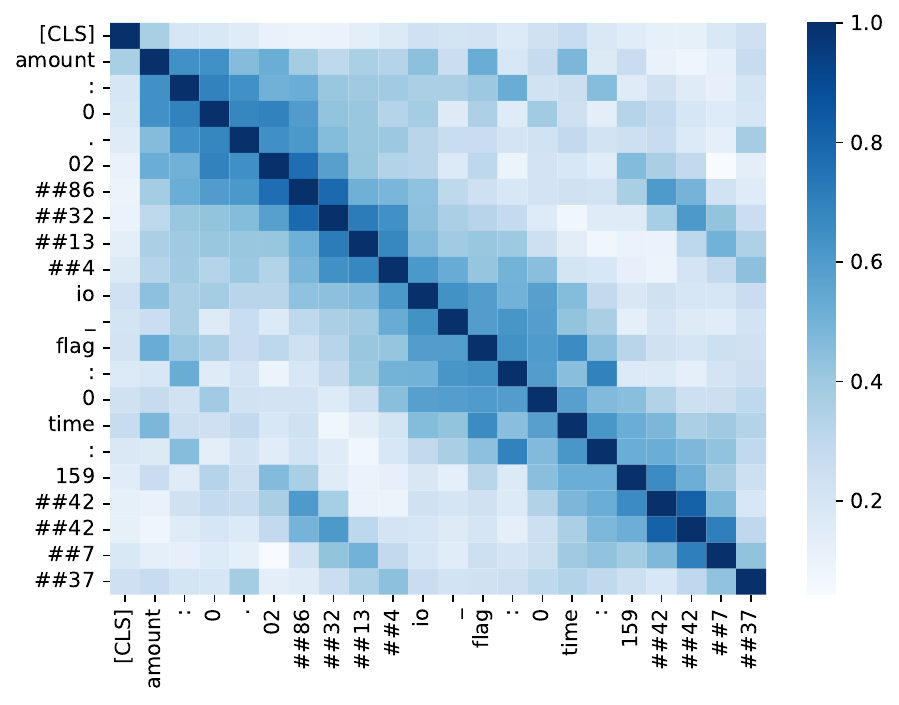}
    \end{minipage}

    \caption{Self-similarity Matrix Visualization: (1) BERT, (2) RoBERTa, (3) ALBERT, (4) ELECTRA, (5) VGCN-BERT, and (6) TxCLM.}
    \label{fig:similarity visualization}
\end{figure*}

\begin{table}[t]
\centering
\caption{Graph sampling statistics and inference efficiency for different methods. Average number of vertices and edges sampled in different layers (All the numbers are in
thousands, lower is better). Last columns show iterations (mini-batches) per second (it/s), higher is better.}
\setlength{\tabcolsep}{1.2mm}
\renewcommand{\arraystretch}{1.05}
\normalsize
\begin{tabular}{c|cccccc|c}
\noalign{\hrule height 1.5pt}
\textbf{Method} & $|\mathbf{V}^{3}|$ & $|\mathbf{E}^{2}|$ & $|\mathbf{V}^{2}|$ & $|\mathbf{E}^{1}|$ & $|\mathbf{V}^{1}|$ & $|\mathbf{E}^{0}|$ & \textbf{it/s} \\ \hline
PLADIES       & \underline{17.2} & 323.0 & 11.0 & 98.3  & 3.1  & 3.9  & 4.9  \\
LADIES        & 17.3 & 305.0 & 11.0 & 95.9  & 3.1  & 3.9  & 3.5  \\
NS            & 25.1 &  \underline{53.7} & \underline{10.7} & \underline{11.9}  & \underline{2.0}  & \underline{2.7}  & \underline{20.2} \\ \hline
LABOR (Ours)  & \textbf{17.6} &  \textbf{34.9} & \textbf{8.3}  & \textbf{11.8}  & \textbf{2.0}  & \textbf{2.7}  & \textbf{18.9} \\
\noalign{\hrule height 1.5pt}
\end{tabular}
\label{tab:graph_sampling_comparison}
\end{table}

To further analyze the effectiveness and advantages of our TxCLM$+$MAGAE+CAFN architecture, we designed additional experiments by replacing various components of our model architecture with alternative baseline methods. Specifically, we replaced the language model and graph model components of our architecture with other baseline methods to verify the improvements in feature extraction and fusion achieved by our method. The CAFN component was kept unchanged, and we fused the features extracted by MAGAE with semantic features obtained from BERT, RoBERTa\cite{roberta}, ALBERT\cite{albert}, ELECTRA\cite{electra}, and VGCN-BERT\cite{vgcn_bert}. Similarly, the interaction features from TxCLM were fused with the features obtained from GCN, GAT, GSAGE, GAE, and GATE. These different model architectures were applied to the B4E dataset, and the experimental results are shown in Tables \ref{tab:repalce_MAGAE} and \ref{tab:replace_TxCLM}.

\begin{figure}[tbp]
\includegraphics[width=0.45\textwidth]{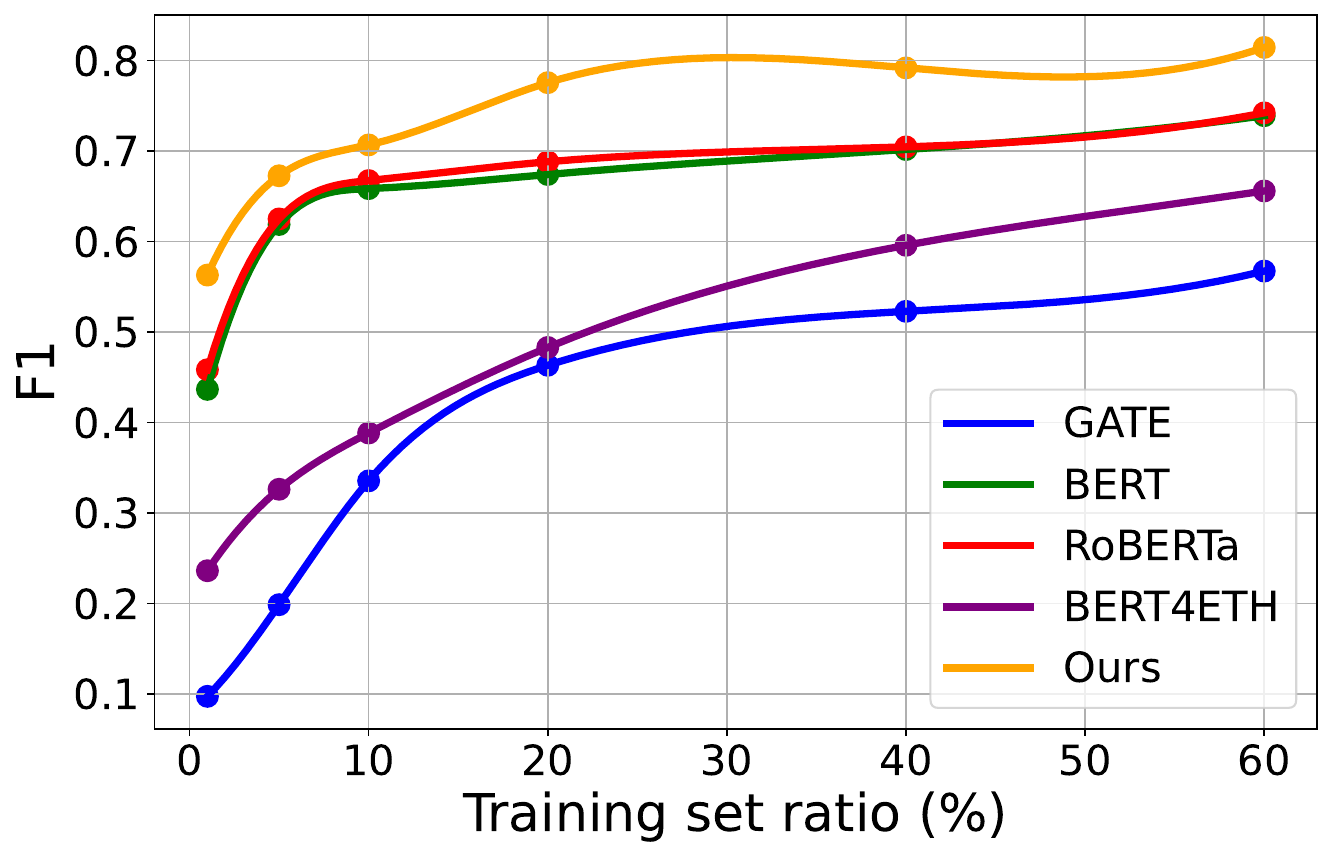}
    \caption{Performance comparison of the respective supervised methods and LMAE4Eth with different sizes of pre-training data on B4E dataset.}
    \label{fig:different pre-training data}
\end{figure}

When the TxCLM component was kept unchanged and the graph model was replaced, our model still showed significant advantages over the TxCLM+Baseline architectures. Specifically, when combining TxCLM with the best-performing supervised method, GAT, LMAE4Eth led by 3.00\% in F1 score and 2.38\% in Balanced Accuracy (BACC). Among the baseline self-supervised methods, the GATE+TxCLM combination performed the best, but still lagged behind our model by 1.02\% in F1 score and 0.94\% in BACC. The node-level reconstruction method in MAGAE provided a greater improvement in fraudulent node classification compared to other graph representation learning methods.

When the MAGAE component was kept unchanged and the language model was replaced, LMAE4Eth still outperformed the MAGAE+Baseline architectures by 2.13\% to 4.25\% in F1 score and 2.02\% to 3.55\% in BACC. Clearly, the token-aware contrastive learning in TxCLM played an important role in these improvements. To analyze the differences in token representations learned by TxCLM and other BERT models, we calculated the average self-similarity of a sequence \( x = [x_1, x_2, ..., x_n] \):
\begin{equation}
\setlength{\abovedisplayskip}{3pt}
\setlength{\belowdisplayskip}{3pt}
    sim(x) = \frac{1}{n(n-1)}\sum_{i=1}^{n}\sum_{j=1,j\neq i}^{n}\textup{cosine}(h_i, h_j),
\end{equation}
where \( h_i \) and \( h_j \) are the token representations of \( x_i \) and \( x_j \) generated by the model. The smaller the value of \( sim(x) \), the lower the similarity between token representations in sequence \( x \), which means the token representations are more discriminative. We randomly selected a transaction record from an account, visualized the self-similarity matrix generated by TxCLM and other BERT models, and observed that darker colors correspond to higher self-similarity scores. As shown in Figure \ref{fig:similarity visualization}, TxCLM generates more differentiated representations for words in the transaction text, which facilitates the model in learning finer-grained semantic representations of the transaction records.

To assess the practical advantage of the layer‑neighbor sampler (LABOR) adopted in MAGAE, we compared it with three representative stochastic samplers—naive node‑wise sampling NS (GraphSAGE)\cite{gsage}, Ladies\cite{zou2019ladies}, and PLadies\cite{labor}—under identical hyper‑parameter settings (fanout {$10{:}10{:}10$}, three‑layer GCN, Adam optimiser with learning rate $0.001$).  All experiments were conducted on the full MulDiGraph dataset.  Table\ref{tab:graph_sampling_comparison} reports, for each sampler, the average number of vertices and edges drawn per layer together with the resulting mini‑batches processed per second.

As shown in Table\ref{tab:graph_sampling_comparison}, LABOR selects only \(8.3\text{k}\) vertices in the second hidden layer and \(34.9\text{k}\) edges in the highest aggregated layer, reducing the edge budget of PLadies and Ladies by an order of magnitude (323.0k and 305.0k, respectively).  Compared with NS, LABOR samples a similar number of shallow‑layer vertices (\(|V^{1}|{=}\,2.0\text{k}\)) but avoids the substantial oversampling at the input layer (\(|V^{3}|{=}\,17.6\text{k}\) vs.\ 25.1 k for NS), yielding a markedly smaller memory footprint. Owing to the reduced edge set, LABOR attains 18.9it/s, nearly a \(\!\times3.8\) speed‑up over PLadies and a \(\!\times5.4\) speed‑up over Ladies, while remaining close to the raw NS throughput.  In practice, this trade‑off—substantially fewer sampled edges with only a marginal throughput drop—proved to be the favourable for training graph model on large‑scale Ethereum graphs. Overall, the statistics confirm that LABOR offers a balanced compromise between sampling sparsity and runtime efficiency.

\subsection{Robustness and Generalization Ability}

The anonymity of accounts in Ethereum makes it difficult to obtain labels, and the vast number of transactions and dynamic changes within the transaction network pose significant challenges for the model's few-shot learning ability and robustness. In this section, we train four self-supervised baseline methods and our model using different portions of the B4E dataset, and fine-tune them with the same data to compare their performance when trained on limited data. Specifically, we experiment with training sizes including 1\% (5,973 samples), 5\% (29,863 samples), 10\% (59,726 samples), 20\% (119,452 samples), 40\% (238,903 samples), and 60\% (358,355 samples). As shown in figure \ref{fig:different pre-training data}, the results on the B4E dataset consistently demonstrate that our method LMAE4Eth outperforms other self-supervised methods across all training data proportions. Notably, even with limited training data, our model shows exceptional proficiency. On the smallest training subset (only 5,973 account samples, 1\%), LMAE4Eth achieved an F1 score of 56.28\%, significantly outperforming the baselines, which range from 9.43\% to 45.28\%. While the performance of the baseline models gradually improves as the number of training samples increases, and the gap with our method narrows, LMAE4Eth still maintains a lead of more than 7\% in F1 score.

\begin{figure*}[t]
    \centering
    \begin{minipage}[b]{0.245\textwidth}
        \centering
        \includegraphics[width=\textwidth]
        {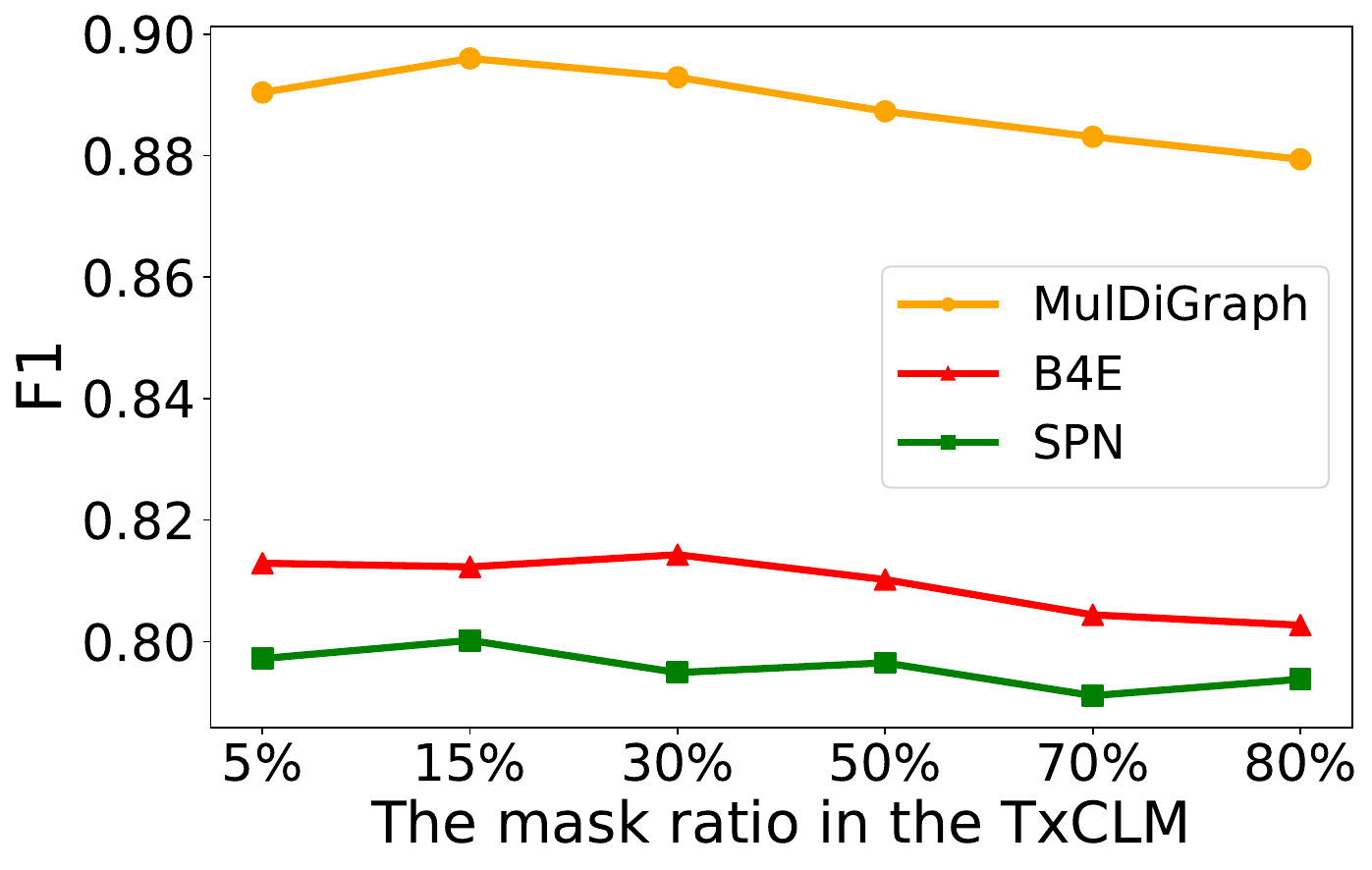}
    \end{minipage}
    \begin{minipage}[b]{0.245\textwidth}
        \centering
        \includegraphics[width=\textwidth]{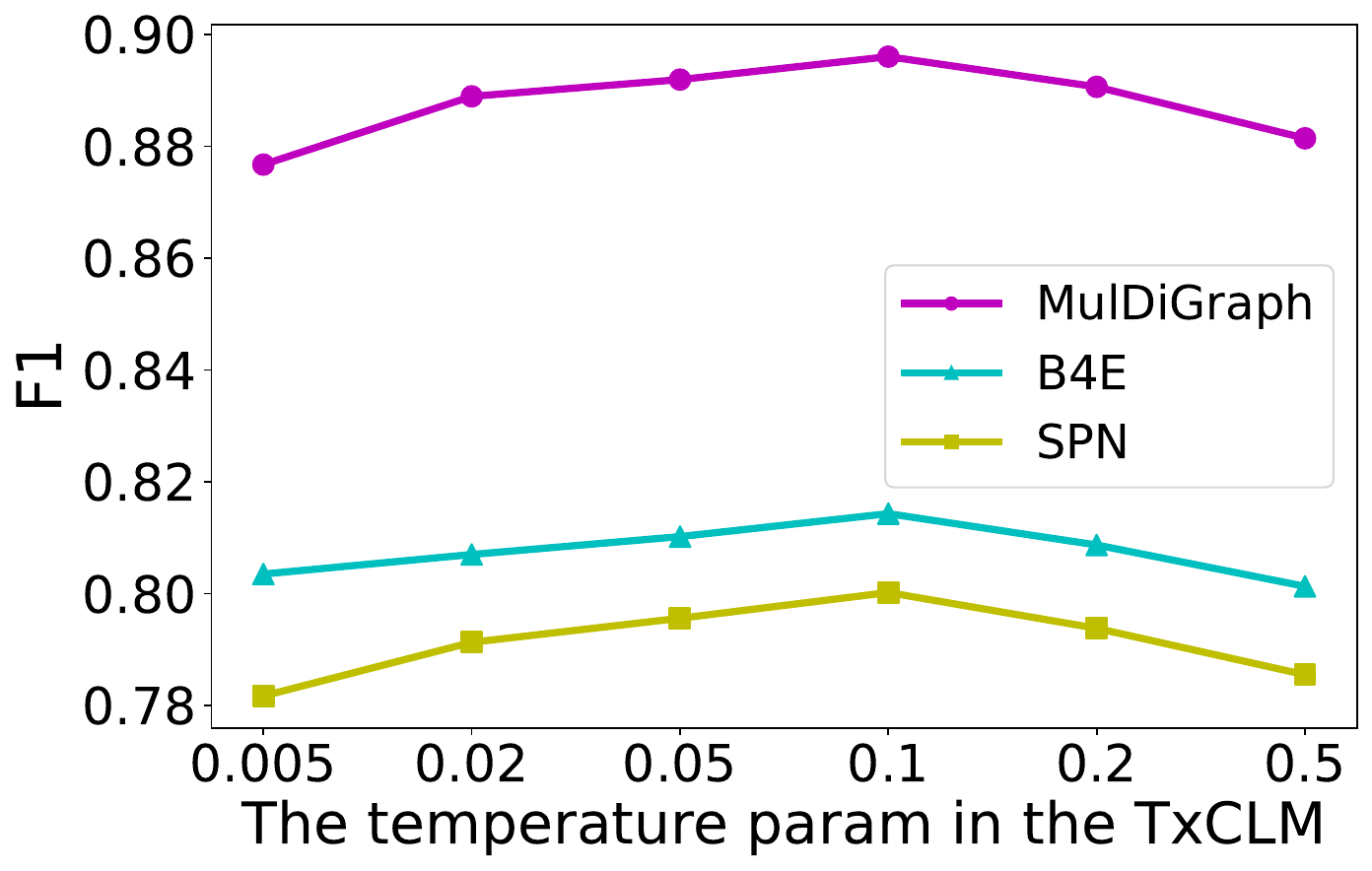}
    \end{minipage}
    \begin{minipage}[b]{0.245\textwidth}
        \centering
        \includegraphics[width=\textwidth]{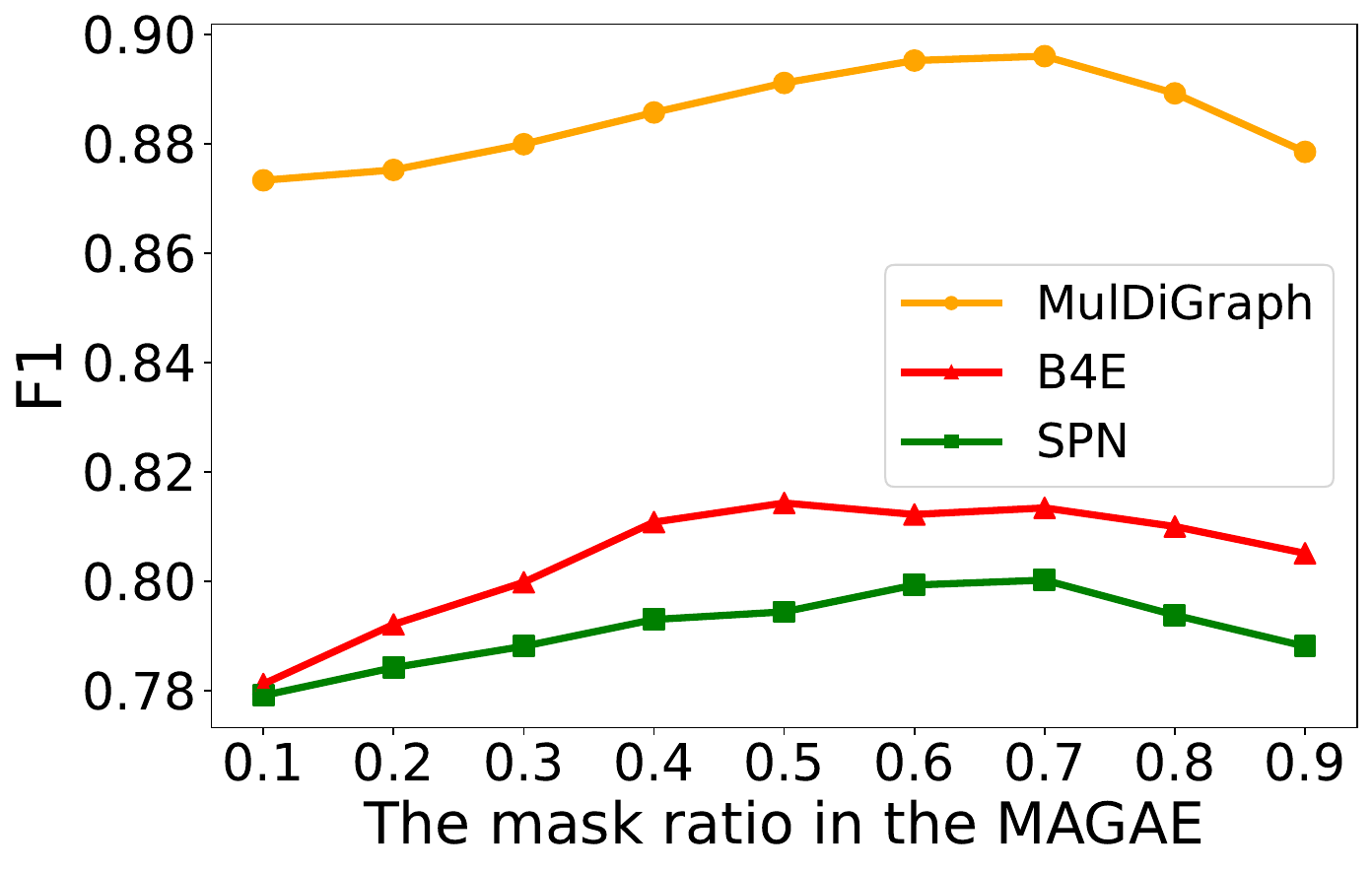}
    \end{minipage}
    \begin{minipage}[b]{0.245\textwidth}
        \centering
        \includegraphics[width=\textwidth]{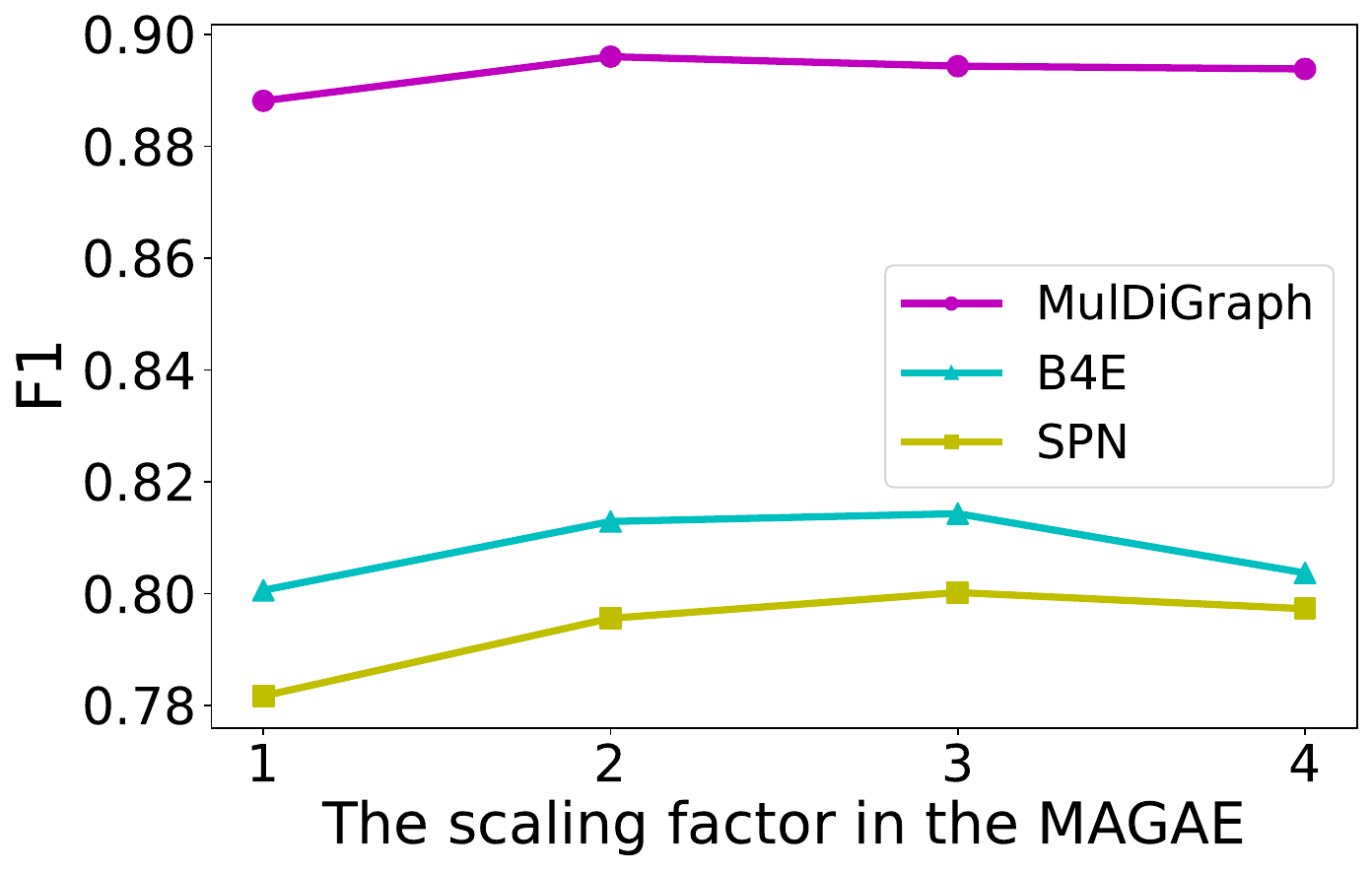}
    \end{minipage}
    \caption{Performance of LMAE4Eth under various critical hyperparameter configurations on three datasets: 
    (a) Mask ratio in the language model, 
    (b) Temperature parameter, 
    (c) Mask ratio  and 
    (d) Scaling factor in the MAGAE.}
    \label{fig:Hyperparameter}
\end{figure*}

\begin{table}[htbp]
\caption{Comparison of F1-score and Balanced Accuracy (BAcc) of Our Method and Baselines Under Isolated Connected-component Splits on Three Datasets.}
\renewcommand{\arraystretch}{1.1}
\setlength{\tabcolsep}{4pt}
\small
\centering
\begin{tabular}{c|cc|cc|cc}
\noalign{\hrule height 1.5pt}
\multirow{2}[0]{*}{Method} & \multicolumn{2}{c|}{D1} & \multicolumn{2}{c|}{D2} & \multicolumn{2}{c}{D3} \\ \cline{2-7}
& \textbf{F1} & \textbf{BAcc} & \textbf{F1} & \textbf{BAcc} & \textbf{F1} & \textbf{BAcc} \\ \hline
GrabPhisher  & 0.7538 & 0.8209 & 0.6641 & 0.7482 & \underline{\textbf{0.7427}} & \underline{\textbf{0.8112}} \\
CATALOG      & 0.7225 & 0.7905 & 0.7117 & \underline{\textbf{0.7827}} & 0.7036 & 0.7768 \\
GraphMLP     & 0.6048 & 0.7078 & 0.3050 & 0.5794 & 0.5385 & 0.6426 \\
GRAND(N)     & 0.6612 & 0.7461 & 0.3847 & 0.6004 & 0.5864 & 0.7008 \\
GRAND        & 0.6285 & 0.7230 & 0.4126 & 0.6047 & 0.5326 & 0.6661 \\
T5           & \underline{\textbf{0.7947}} & \underline{\textbf{0.8478}} & 0.6335 & 0.7268 & 0.7325 & 0.8003 \\
Longformer   & 0.7543 & 0.8178 & \underline{\textbf{0.6912}} & 0.7671 & 0.7277 & 0.7938 \\ \hline
\textbf{Ours} & \textbf{0.8996} & \textbf{0.9262} & \textbf{0.8216} & \textbf{0.8688} & \textbf{0.7895} & \textbf{0.8410} \\
\noalign{\hrule height 1.5pt}
\end{tabular}
\renewcommand{\arraystretch}{1}
\label{tab:component_split_f1_bacc}
\end{table}

To further validate the robustness and generalization ability of our method, we conducted an additional controlled experiment using strictly isolated connected components graph. For each dataset, we selected all phishing-labeled accounts and randomly sampled twice as many benign accounts to construct a label-imbalanced transaction graph. Based on complete transaction records, we extracted all connected components and partitioned them into training, validation, and testing sets following a 7:1:2 ratio, ensuring no edge connectivity between these sets. This setup mimics a more challenging inductive learning scenario where models must classify accounts in unseen subgraphs. This procedure yields 2860, 408, and 818 connected components for training, validation, and testing on MulDiGraph (D1), 6762, 966, and 1932 for B4E (D2), and 12417, 1479, and 2958 for SPN (D3), respectively. We evaluated our method and several strong baselines under this setting. As shown in Table~\ref{tab:component_split_f1_bacc}, our model consistently outperforms all baselines across all datasets, achieving approximately 10\% F1 gain over the best-performing baseline on D1 and D2, and a 5\% gain on D3.

Notably, graph-based methods such as GraphMLP and GRAND experience considerable performance degradation, highlighting their limited generalization when structural overlap is removed. In contrast, Transformer-based models like CATALOG, T5, and Longformer retain relatively stable performance. Despite partially utilizing graph structure, our model maintains superior performance. We attribute this to: (1) the semantic modeling capacity of the language component, which captures informative transactional patterns independent of structural proximity, and (2) the higher-level integration of graph structure via attention mechanisms, which reduces sensitivity to local neighborhood dependencies.

\subsection{Hyperparameter Studies}

\begin{figure}[t]
\includegraphics[width=0.45\textwidth]{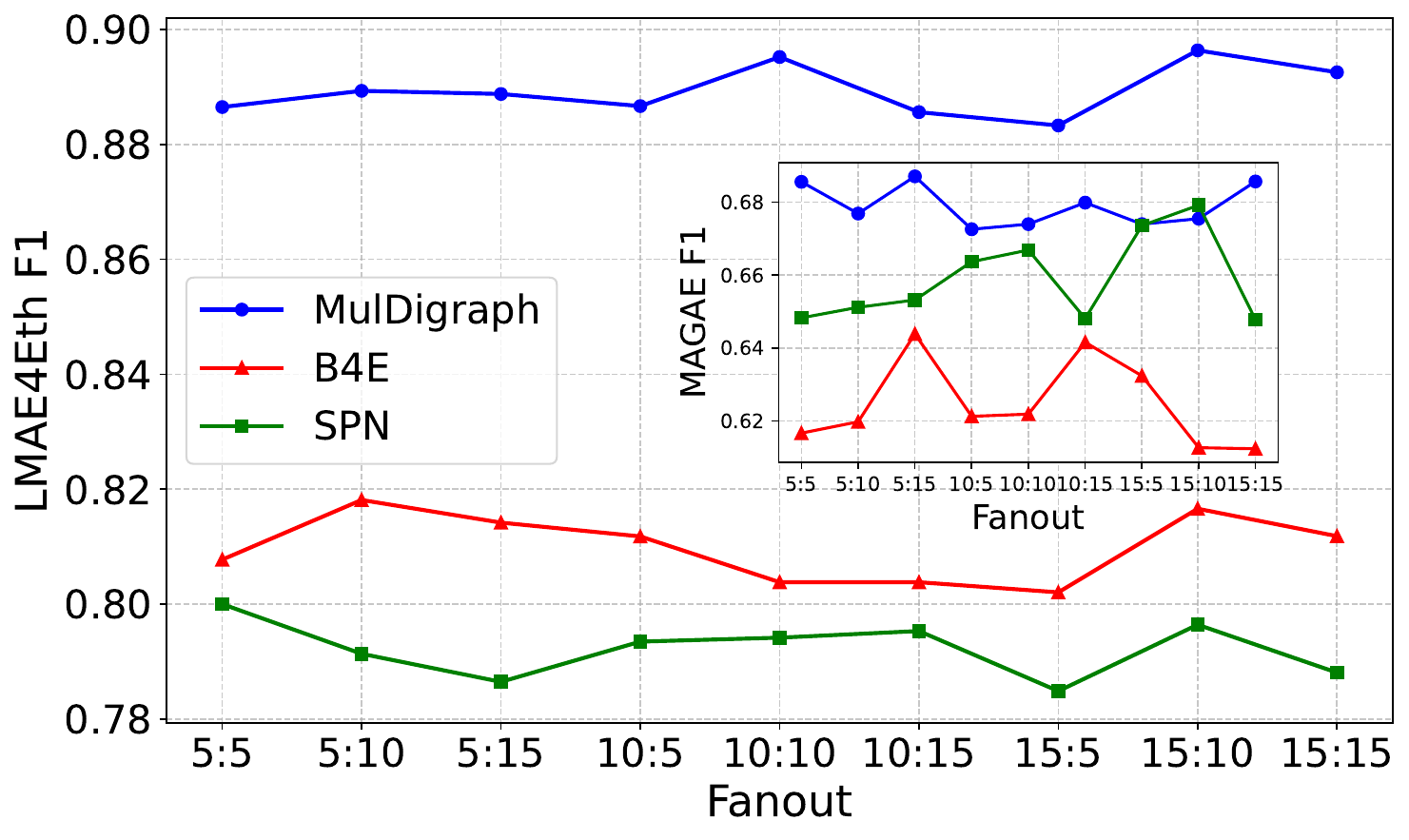}
    \caption{Performance of LMAE4Eth and MAGAE alone (inset figure) under different predetermined numbers of nodes in layer-neighbor sampling (Fanout) across three datasets.}
    \label{fig:sampler_nodes}
\end{figure}


To demonstrate the effects of different parameter settings, we conducted experiments to evaluate the performance of our proposed LMAE4Eth framework under various key hyperparameter configurations. These configurations include the masking ratio and temperature parameter in TxCLM, as well as the node masking ratio and scaling factor in MAGAE. When analyzing the effect of a specific hyperparameter, all other parameters were kept at their default values. The results on the three datasets are presented in the figures, and the observations are summarized below to analyze the impact of different hyperparameters:

We adjusted the masking ratio in the language model across \{5\%, 15\%, 30\%, 50\%, 70\%, 80\%\}. The LMAE4Eth achieved optimal performance on MulDiGraph and SPN with a 15\% masking ratio, while the best performance on B4E was obtained with a 30\% masking ratio. However, as the masking ratio increased further, prediction accuracy declined slightly. Higher masking ratios posed challenges for the model to capture transactional semantic features, particularly in Ethereum transaction texts that exhibit high homogeneity.

In TxCLM, the temperature parameter \( \tau \) is used to control the sensitivity of the similarity scores in the loss function, influencing the model’s sensitivity to differences between samples. We adjusted \( \tau \) within the range \{0.005, 0.02, 0.05, 0.1, 0.2, 0.5\}, and the results indicated that maintaining a moderately high temperature parameter (e.g., 0.1) enabled the model to effectively capture subtle distinctions in transaction tokens.

For MAGAE, we explored the impact of the node masking ratio across \{0.1, 0.2, 0.3, 0.4, 0.5, 0.6, 0.7, 0.8, 0.9\}. The results on all three datasets revealed that a low masking ratio failed to provide sufficient challenge for the reconstruction task, making it difficult for the model to learn meaningful information. Conversely, higher masking ratios (between 0.5 and 0.8) significantly enhanced model performance.

The scaling factor \( \gamma \) in MAGAE adjusts the sensitivity of the Scaled Cosine Error to address the imbalance between easy and hard samples during the feature reconstruction process. For Ethereum fraud detection, which is inherently an imbalanced classification task, our experiments showed that higher scaling factors within the range \{1, 2, 3, 4\} allowed the model to focus more on difficult samples during training, improving the autoencoder’s ability to reconstruct abstract interaction features.

The fanout parameter determines the number of nodes sampled from the neighbors of a node when aggregating at each layer of GNN. We explored the impact of sampling the two-layer neighborhood with the number of \(\{|V^1|:|V^2|\} = \{5\!:\!5,\, 5\!:\!10,\, 5\!:\!15,\, 10\!:\!5,\, 10\!:\!10,\, 10\!:\!15,\, 15\!:\!5,\, 15\!:\!10,\, 15\!:\!15\}\) on MAGAE and the overall model. As shown in the figure~\ref{fig:sampler_nodes}, we found that the best fanout on different datasets is not fixed, and its impact on the overall LMAE4Eth model is limited. In particular, we found that the graph model does not necessarily achieve the best performance when the fanout is high. For example, the graph model performs worst when the two-layer neighbors are sampled at \(15\!:\!15\) on the B4E dataset. We speculate that this is because although a larger fanout can sample more neighbors, it may also introduce more noise that reduces the model's ability to perceive phishing nodes.

\section{Limitations and Future Work}
\label{sec:limitations}
While LMAE4Eth demonstrates promising results across multiple dataset, several aspects offer room for future enhancement. Similar to most previous studies, the current framework is designed for batch-mode analysis, operating under the assumption that accounts have accumulated sufficient transaction history and that an approximate snapshot of the account–transaction graph is accessible during training. This design, while effective in retrospective settings, may limit applicability in scenarios requiring real-time inference or the handling of newly created accounts. Although the use of LABOR sampling substantially improves computational efficiency, the ability to incrementally update model states remains a direction worth pursuing. In addition, all experiments are conducted on publicly released Ethereum datasets. While these datasets are derived from real-world on-chain interactions and are widely adopted in existing studies, they do not fully capture evolving behaviors across different platforms and real-time updated on-chain transactions. 

As future work, we aim to incorporate streaming and incremental learning techniques to support low-latency updates and improve coverage of low-activity accounts. We also plan to explore collaboration with blockchain service providers to access more diverse and timely datasets, enabling deployment in dynamic or multi-chain environments.

\section{Conclusion}

In this paper, we introduce LMAE4Eth, a novel self-supervised approach that integrates transaction language models with graph-based methods to capture both semantic and structural features of transaction data in Ethereum.    Our work represents the first attempt to utilize language models to address the challenge of unclear transaction semantics, and we are pioneers in modeling Ethereum account features from a dual-perspective view.    We propose a token-aware contrastive learning enhanced transaction language model and introduce node-level reconstruction using a masked graph autoencoder, along with layer-neighbor sampling strategies, into the Ethereum transaction network.    Additionally, we present a cross-attention based fusion network that successfully integrates semantic and interactional perspectives features.    Our approach has demonstrated significant improvements, achieving performance gains of 6\% to 10\% across three datasets compared to current state-of-the-art methods.    These empirical results provide strong evidence for the effectiveness of our proposed methodology, highlighting the potential of combining linguistic, semantic, and structural analysis in blockchain analytics and fraud detection.

\footnotesize
\bibliographystyle{IEEEtran}
\bibliography{tifs-template}

\end{sloppypar}

\end{document}